\documentclass[12pt]{article}
\def\one{
\setlength{\unitlength}{0.45cm}
\begin{picture}(0.55,0.5)
\put(0,0){\line(1,0){0.4}}
\put(0,.4){\line(1,0){0.4}}
\multiput(0,0)(.4,0){2}{\line(0,1){.4}}
\end{picture}}
\def\twohor{
\setlength{\unitlength}{0.45cm}
\begin{picture}(1.1,0.5)
\put(0,0){\line(1,0){0.8}}
\put(0,.4){\line(1,0){0.8}}
\multiput(0,0)(.4,0){3}{\line(0,1){.4}}
\end{picture}}
\def\twover{
\setlength{\unitlength}{0.45cm}
\begin{picture}(0.55,0.5)
\put(0,0){\line(1,0){0.4}}
\put(0,.4){\line(1,0){0.4}}
\put(0,-.4){\line(1,0){0.4}}
\multiput(0,0)(.4,0){2}{\line(0,0){.4}}
\multiput(0,0)(.4,0){2}{\line(0,-1){.4}}
\end{picture}}
\def\threehor{
\setlength{\unitlength}{0.45cm}
\begin{picture}(1.5,0.5)
\put(0,0){\line(1,0){1.2}}
\put(0,.4){\line(1,0){1.2}}
\multiput(0,0)(.4,0){4}{\line(0,1){.4}}
\end{picture}}
\def\threever{
\setlength{\unitlength}{0.45cm}
\begin{picture}(0.6,0.5)
\put(0,0){\line(1,0){0.4}}
\put(0,.4){\line(1,0){0.4}}
\put(0,-.4){\line(1,0){0.4}}
\put(0,-.8){\line(1,0){0.4}}
\multiput(0,0)(.4,0){2}{\line(0,0){.4}}
\multiput(0,0)(.4,0){2}{\line(0,-1){.4}}
\multiput(0,0)(.4,0){2}{\line(0,-2){.8}}
\end{picture}}
\def\mixed{
\setlength{\unitlength}{0.45cm}
\begin{picture}(1,0.5)
\put(0,0){\line(1,0){0.8}}
\put(0,.4){\line(1,0){0.8}}
\put(0,-.4){\line(1,0){0.4}}
\multiput(0,0)(.4,0){3}{\line(0,1){.4}}
\multiput(0,0)(.4,0){2}{\line(0,-1){.4}}
\end{picture}}

\usepackage{epsf}
\usepackage{cite}
\def\one{
\setlength{\unitlength}{0.45cm}
\begin{picture}(0.55,0.5)
\put(0,0){\line(1,0){0.4}}
\put(0,.4){\line(1,0){0.4}}
\multiput(0,0)(.4,0){2}{\line(0,1){.4}}
\end{picture}}
\def\twohor{
\setlength{\unitlength}{0.45cm}
\begin{picture}(1.1,0.5)
\put(0,0){\line(1,0){0.8}}
\put(0,.4){\line(1,0){0.8}}
\multiput(0,0)(.4,0){3}{\line(0,1){.4}}
\end{picture}}
\def\twover{
\setlength{\unitlength}{0.45cm}
\begin{picture}(0.55,0.5)
\put(0,0){\line(1,0){0.4}}
\put(0,.4){\line(1,0){0.4}}
\put(0,-.4){\line(1,0){0.4}}
\multiput(0,0)(.4,0){2}{\line(0,0){.4}}
\multiput(0,0)(.4,0){2}{\line(0,-1){.4}}
\end{picture}}
\def\threehor{
\setlength{\unitlength}{0.45cm}
\begin{picture}(1.5,0.5)
\put(0,0){\line(1,0){1.2}}
\put(0,.4){\line(1,0){1.2}}
\multiput(0,0)(.4,0){4}{\line(0,1){.4}}
\end{picture}}
\def\threever{
\setlength{\unitlength}{0.45cm}
\begin{picture}(0.6,0.5)
\put(0,0){\line(1,0){0.4}}
\put(0,.4){\line(1,0){0.4}}
\put(0,-.4){\line(1,0){0.4}}
\put(0,-.8){\line(1,0){0.4}}
\multiput(0,0)(.4,0){2}{\line(0,0){.4}}
\multiput(0,0)(.4,0){2}{\line(0,-1){.4}}
\multiput(0,0)(.4,0){2}{\line(0,-2){.8}}
\end{picture}}
\def\mixed{
\setlength{\unitlength}{0.45cm}
\begin{picture}(1,0.5)
\put(0,0){\line(1,0){0.8}}
\put(0,.4){\line(1,0){0.8}}
\put(0,-.4){\line(1,0){0.4}}
\multiput(0,0)(.4,0){3}{\line(0,1){.4}}
\multiput(0,0)(.4,0){2}{\line(0,-1){.4}}
\end{picture}}

\catcode`@=11 \@addtoreset{equation}{section} \catcode`@=12

\def\p2{\frac{p}{2}}
\def\tp2{\frac{3p}{2}}

\begin{document}

\setlength \arraycolsep{2pt}

\begin{titlepage}
\vfill
\begin{center}
{\Large \bf Composite Invariants and Unoriented   
Topological String Amplitudes}\\[1cm] 
Chandrima Paul\footnote{E-mail:chandrima@phy.iitb.ac.in}, 
Pravina Borhade \footnote{E-mail: borhadepravina@gmail.com}, 
P. Ramadevi\footnote{Email: ramadevi@phy.iitb.ac.in}\\
{\em Department of Physics, \\Indian Institute of Technology Bombay,\\
Mumbai 400 076, India\\[10pt]}
\end{center}
%\vspace{2cm}
\vfill
\begin{abstract}
Sinha and Vafa \cite {sinha} had conjectured that the $SO$ Chern-Simons 
gauge theory on $S^3$ must be dual to the closed $A$-model 
topological string on the orientifold of a resolved conifold. 
Though the Chern-Simons free energy could be rewritten in terms 
of the topological string amplitudes providing evidence
for the conjecture, we needed a novel idea in the 
context of Wilson loop observables to extract cross-cap 
$c=0,1,2$ topological amplitudes. Recent paper of Marino \cite{mar9} 
based on the work of Morton and Ryder\cite{mor} has clearly 
shown that the composite representation placed on the knots 
and links plays a crucial role to rewrite the topological 
string cross-cap $c=0$ amplitude. This enables extracting the 
unoriented cross-cap $c=2$ topological  amplitude. In this paper, 
we have explicitly worked out the composite invariants 
for some framed knots and links carrying 
composite representations in $U(N)$ Chern-Simons theory. 
We have verified generalised Rudolph's theorem,
which relates composite  invariants to the invariants in 
$SO(N)$ Chern-Simons theory, and also verified Marino's conjectures
on the integrality properties of the topological
string amplitudes. For some framed knots and links, we have 
tabulated the BPS integer invariants for cross-cap $c=0$, $c=1$ and 
$c=2$ giving the open-string topological amplitude 
on the orientifold of the resolved conifold.
\end{abstract}
\vfill
\end{titlepage}
\section{Introduction}
We have seen interesting developments in the open string and 
closed string dualities during the last 12 years starting
from the celebrated work of Maldacena \cite{malda}. 
Gopakumar and Vafa \cite {gv1,gv2,gv3} conjectured  open-closed 
duality in the topological string context. 
Gopakumar-Vafa conjecture states that the $A$-model
open topological string theory on the deformed conifold,
equivalent to  the Chern-Simons gauge theory on $S^3$ \cite {wittencs}, 
is dual to the closed string theory on a resolved conifold. 

In ref. \cite {gv1}, it was shown that the free-energy expansion 
of $U(N)$ Chern-Simons field theory on $S^3$ at large $N$  resembles 
$A$-model topological string theory amplitudes on the resolved conifold.
This provided an evidence for the conjecture. Another piece of
evidence at the level of observables was shown by Ooguri and 
Vafa \cite{ov} for the simplest Wilson loop observable (simple 
circle also called unknot) in Chern-Simons theory on $S^3$. 
In particular, Ooguri-Vafa considered the expectation value of a 
scalar operator ${\cal Z}_{\cal H}(v)$ in the topological 
string theory corresponding to the simple circle in submanifold 
$S^3$ of the deformed conifold and showed its  form in the resolved 
conifold background. From these results for unknot,
Ooguri-Vafa conjectured on the form for  
${\cal Z}_{\cal H}(v)$ for any knot or link in $S^3$.
For completeness and simplicity, we briefly present the form for knots:  
\begin{eqnarray}
{\cal F}_{\cal H}(v)&=&\ln {\cal Z}_{\cal H}(v)=\ln\{
\sum_R {\cal H}_R[{\cal K}] s_R(v)\} 
~=\sum_{R,d} f_R(q^d,\lambda^d) s_R(v^d)~\\
{\rm where}~f_R(q,\lambda)&=& {1 \over (q^{1/2}-q^{-1/2})}
\sum_{Q,s}N_{R,Q,s} \lambda^Q q^s \label {unrfm}
\end{eqnarray}
Here ${\cal H}_R({\cal K})$ are the $U(N)$ Chern-Simons invariants
for a knot ${\cal K}$ in $S^3$ carrying representation $R$ and $s_R(v)$
are the Schur polynomials in variable $v$ which represent
$U(N)$ holonomy of the knot ${\cal K}$ in the Lagragian submanifold
${\cal N}$ which intersects $S^3$ along the knot. ${\cal F}_{\cal H}(v)$
denotes the free-energy of the topological open-string partition function
on the resolved conifold and $f_R(q,\lambda)$ are the
$U(N)$ reformulated invariants. The conjecture states that the 
reformulated invariant must have the form (\ref {unrfm}) where 
$N_{R,Q,s}$ are integer coefficients.

Labastida-Marino \cite{lm} used group-theoretic 
techniques to rewrite the expectation value of the topological operators 
in terms of link invariants in $U(N)$ Chern-Simons field theory
on $S^3$.  This group theoretic approach enabled verification of 
Ooguri-Vafa conjecture for many non-trivial knots\cite{lm, taps, laba, mari1}. 
Conversely, the Ooguri-Vafa conjecture led to a reformulation of 
Chern-Simons field theory invariants for knots and links giving new
polynomial invariants(\ref {unrfm}). The integer coefficients of these
new polynomial invariants have topological meaning accounting
for BPS states in the string theory. The challenge still  
remains in obtaining such integers for non-trivial knots
and links within topological string theory.

Another challenging question is to attempt 
similar duality conjectures between Chern-Simons gauge theories 
on three-manifolds other than $S^3$ and closed string theories. 
Invoking Gopakumar-Vafa conjecture and Ooguri-Vafa conjecture,  
it was possible to explicitly write the $U(N)$ Chern-Simons 
free-energy expansion at large $N$ as a closed string 
theoretic expansion\cite{ramprav}. Surprisingly, the  expansion 
resembled partition function of a closed string theory on a 
Calabi-Yau background with one kahler parameter. 
Unfortunately,  the Chern-Simons free-energy expansion 
for other three-manifolds are not equivalent to the `t Hooft 
large $N$ perturbative expansion around a classical solution \cite {thoo}.
In order to predict new duality conjectures, we need to extract 
the perturbative expansion around a classical solution from the 
free-energy.

For orbifolds of $S^3$, which gives Lens space ${\cal L}[p,1] \equiv S^3/Z_p$,
it is believed that the Chern-Simons theory is dual to the
$A$-model closed string theory on $A_{p-1}$ 
fibred over $P^1$ Calabi-Yau background.  It was Marino \cite {marin} who
showed that the perturbative Chern-Simons theory on 
Lens space ${\cal L}[p,1]$ can be given a  matrix model description. 
Also, hermitian matrix model description of $B$-model topological
strings \cite {dijk} was shown to be equivalent to Marino's matrix model
using mirror symmetry\cite{akmv}. It is still a challenging open problem 
to look for dual closed string description corresponding to $U(N)$ Chern-Simons
theory on other three-manifolds.

The extension of these duality conjectures for other gauge 
groups like $SO(N)$ and $Sp(N)$ have also been studied.
In particular, the free-energy expansion $F^{(SO)}_{(CS)}[S^3]$ of
the Chern-Simons theory on $S^3$ based on $SO$ gauge group was 
shown to be dual to $A$-model closed string theory on a orientifold 
of the resolved conifold background \cite {sinha}. In particular, 
the string partition function $Z$ for these orientifolding action 
must have two contributions: 
\begin{equation}
F^{(SO)}_{(CS)}[S^3]~=Z~=~{1 \over 2}Z^{or} + Z^{(unor)} \label {sinh}
\end{equation}
where $Z^{(or)}$ is the untwisted contribution and $Z^{(unor)}$
is the twisted sector contribution. The untwisted contribution
exactly matches the $U(N)$ Chern-Simons free energy on $S^3$. 
Using the topological vertex as a tool, Bouchard et al \cite {vinc1,vinc2} 
have determined unoriented closed string amplitude and unoriented 
open topological string amplitudes for a few orientifold toric geometry 
with or without $D$-branes. 

In Ref.\cite {prav}, the generalisation of Ooguri-Vafa conjecture 
for observables involving $SO(N)$ holonomy, different 
from the works of Bouchard et al \cite{vinc1,vinc2}, was studied.
Similar to the $U(N)$ result (\ref {unrfm}), the coefficients of 
$SO(N)$ reformulated invariants are indeed integers.  
%The reformulated invariants for knots in 
%standard framing obeyed the conjecture of 
%Bouchard-Florea-Marino\cite {vinc2} giving the integer coefficients 
%corresponding to cross-cap $c=1$ unoriented open-string amplitude. 
%We also generalised the Bouchard-Florea-Marino conjecture for any 
%$r$-component framed links and have verified for few examples of 
%framed knots and two-component framed links.

Following Sinha-Vafa conjecture \cite{sinha}, the expectation value of the 
topological string operator (observables) $Z_{\cal G}(v)$ 
where ${\cal G}$ represents $SO(N)$ knot invariants in 
Chern-Simons theory on $S^3$ and $v$ represents the $SO$ holonomy on
the submanifold ${\cal N}$ intersecting $S^3$ along a knot. It is expected
that the free-energy of the open-string partition function on the orientifold
of the resolved conifold must also satisfy a relation 
similar to eqn.(\ref{sinh}): 
\begin{equation}
{\cal F}_{\cal G}(v)= \ln Z_{\cal G}(v)= {1 \over 2} {\cal F}^{(or)}_{\cal R}(v)
+{\cal F}^{(unor)}(v)~. \label {orie1} 
\end{equation}
where ${\cal F}^{(or)}_{\cal R}(v)$ is the oriented or
untwisted sector contribution (also called cross-cap $c=0$)
and the twisted sector term ${\cal F}^{(unor)}(v)$ will have both 
cross-cap $c=1$ and $c=2$ contributions to the open topological
string amplitudes.
It was not clear \cite{vinc1,vinc2} as to how to obtain 
${\cal F}^{(or)}_{\cal R}(v)$ in the orientifold theory 
using $U(N)$ Chern-Simons knot invariants.
As a result, it was not possible to distinguish the topological
amplitudes of cross-cap $c=0$ from 
$c=2$ contribution. However using parity argument
in variable $\sqrt{\lambda}$, the cross-cap $c=1$ topological
amplitudes contribution could be obtained\cite{vinc1,vinc2,prav}. 

From the orientifolding action, Marino \cite{mar9} has indicated 
that there must be a $U(N)$ composite
representation $(R,S)$ placed on the knot in $S^3$ and the
oriented contribution must be rewritable as: 
\begin{equation}
{\cal F}^{(or)}_{\cal R}(v)=\sum_{R,S}{\cal H}_{(R,S)}[{\cal K}] s_R(v) s_S(v)
=\sum_R {\cal R}_R[{\cal K}] s_R(v) \label {orient}
\end{equation}
where $s_R(v)$ and $s_S(v)$ are the Schur polynomials 
corresponding to the $U(N)$ holonomy in two Lagrangian
submanifolds ${\cal N}_{\epsilon}$ and 
${\cal N}_{-\epsilon}$ related by the orientifolding action.
Here $\epsilon$ denotes the deformation parameter of the 
deformed conifold. The oriented invariant ${\cal R}_R[{\cal K}]$
can be obtained from composite invariants ${\cal H}_{(R,S)}[{\cal K}]$ 
using the properties of the Schur polynomials. Though we have so far 
discussed for knots, it is straightforward to generalise 
these arguments for any $r$-component link $L$.

In this paper, we explicitly evaluate the 
composite invariants ${\cal H}_{(R_1,S_1),(R_2,S_2),\ldots (R_r,S_r)}[L]$, 
in $U(N)$ Chern-Simons gauge theory for many framed knots and links $L$
made of $r$ component knots ${\cal K}_{\alpha}$'s carrying
composite representations $(R_{\alpha},S_{\alpha})$ 
using the tools\cite{rama}. These composite invariants
are polynomials in two variables $q,\lambda$. 
We find that the framing factor for the component knots of the 
links carrying composite representation requires a 
slightly modified choice of the $U(1)$ charge so that the 
composite invariants are polynomials in variables
$q$ and $\lambda$.  

Comparing these invariants with $SO(N)$ Chern-Simons invariants 
${\cal G}_{R_1,R_2,\ldots R_r}[L]$\cite{prav} for link
$L$ whose components carry representations $R_{\alpha}$'s which are 
also polynomials in two variables $(q,\lambda)$, 
we have verified the generalised Rudolph's theorem\cite{mor,Rudo}:
\begin{equation}
{1 \over 2} \left[ {\cal H}_{(R,R)}[{\cal K}]+
\{{\cal G}_R[{\cal K}]\}^2 \right]=
f(q)\sum_{n,p} a_{n,p} \lambda^{n \over 2} q^p~, \label {rud}
\end{equation}
for many framed knots ${\cal K}$ carrying $R=\twover, \twohor, \one$. 
Here $f(q)$ is a function of $q$, $a_{n,p}$ are integers. In fact, the 
above relation between $U(N)$ composite invariants and 
$SO(N)$ invariants appears naturally from the integrality properties 
of the topological string amplitudes in the orientifold geometry \cite{mar9}.
Using these composite representation invariants, we  
verified the integrality conjectures of Marino\cite{mar9}
for  framed knots and framed two-component links.  
While submitting this paper, we came across a recent paper\cite {stev}
where Marino's conjectures have been verified for standard framing 
torus knots and torus links which is a special case of our results.

The organisation of the paper is as follows. In
section 2, we present composite framed knot and 
framed two-component link invariants in $U(N)$ Chern-Simons theory.  
In section 3, we briefly review Marino's conjectures
on the reformulated invariants of the framed links 
in the orientifold resolved conifold. 
In section 4, we verify one of Marino's conjectures
and tabulate the $c=0$ BPS integer coefficients
for few examples. In section 5, we obtain
the reformulated invariants corresponding to the
unoriented topological string amplitude. Indeed, these
reformulated invariants also obey the integrality
conjecture of Marino. We have tabulated the 
$c=1$ and $c=2$ BPS integers for some framed knots and framed Hopf link
in section 6. In the concluding section, we summarize the results 
obtained.  In appendix A, we present $U(N)$ composite invariants  for 
some framed knots and framed two-component links for some representations. 
In appendix B, the unoriented reformulated polynomial invariants for 
few non-trivial framed knots and framed links are presented. 

\section{Chern-Simons Gauge theory and Composite Link invariants}
Chern-Simons gauge theory on $S^3$ based on the gauge group $G$ 
is described by the following action:
\begin{equation}
S = {k \over 4 \pi} \int_{S^3} Tr\left (A \wedge dA + {2 \over 3} A \wedge
A \wedge A \right)  
\end{equation}
where $A$ is a gauge connection for compact semi-simple gauge group $G$ and 
$k$ is the coupling constant. The observables in this theory are 
Wilson loop operators: 
\begin{equation}
W_{R_1,R_2, \ldots R_r}[L]~=~ \prod_{\alpha=1}^rTr_{R_{\alpha}} 
U [{\cal K}_{\alpha}]~,
\end{equation}
where $U[{\cal K}_{\alpha}]=P\left[\exp \oint_{{\cal K}_{\alpha}} A\right]$
denotes the holonomy of the gauge field $A$ around the 
component knot ${\cal K}_{\alpha}$
of a $r$-component link $L$  carrying 
representation $R_{\alpha}$. The expectation value of these Wilson loop 
operators are the link invariants:
\begin{equation}
\langle W_{R_1,R_2, \ldots R_r}[L] \rangle(q,\lambda)= {\int[{\cal D}A]e^{iS}
 W_{R_1,R_2,\ldots,R_r}[L] \over \int[{\cal D} A]e^{iS}}~, 
 \label {linki} 
\end{equation}
These link invariants are polynomials in two variables 
\begin{equation}
q=\exp\left({2 \pi i \over k+C_v}\right)~,~ \lambda = q^{N+a}~,
\end{equation}
where $C_v$ is the dual coxeter number of the gauge group $G$
\begin{displaymath}
\begin{array} {lcl}
C_v = \left\{ \begin{array}{ll}
N & {\rm for}~ G=SU(N)\\
N-2 & {\rm for} ~G=SO(N)\\
\end{array} \right. & {\rm and} &
a = \left\{ \begin{array}{ll}
0 & {\rm for}~ G=SU(N)\\
-1 & {\rm for} ~G=SO(N)\\
\end{array} \right.
\end{array}
\end{displaymath}
These link invariants can be computed using the following two 
inputs\cite{rama}:\\
(i) Any link can be drawn as a closure or plat of braids,~\\
(ii) The connection between Chern-Simons theory and the Wess-Zumino 
conformal field theory.\\
We now define some quantities which will be useful later. The quantum
dimension of a representation $R$ with
highest weight $\Lambda_R$ is given by
\begin{equation}
dim_q R=\Pi_{\alpha>0}\frac{[\alpha\cdot (\rho+\Lambda_R)]
}{[\alpha\cdot\rho]}~, \label {quan}
\end{equation}
where $\alpha$'s are the positive roots and $\rho$ is the Weyl vector
equal to the sum of the fundamental weights of the group $G$.
The square bracket refers to the quantum number defined by
\begin{equation}
[x]={\left(q^{x/2}-q^{-x/2} \right) \over \left(q^{1/2}- q^{-1/2} \right)}~.
\label{qno}
\end{equation}
The $SU(N)$ quadratic Casimir for representation $R$ is given by
\begin{equation}
C_R = -\frac{\ell^2}{2N}+\kappa_R =  -\frac{\ell^2}{2N}+
\frac{1}{2} \left ((N+a) \ell + \ell + \sum_i (l_i^2- 2i l_i) \right)~.
\label {casi}
\end{equation}
Our interest is to obtain invariants of framed knots and framed links
carrying representation $R_c \equiv (R,S)$ called composite 
representation in $U(N)$ Chern-Simons gauge theory so that
Marino's conjectures on the topological amplitudes
in the orientifold of resolved conifold geometry can be
verified.
%The works of Morton-Ryder \cite{mor} and   
%Marino\cite{mar9} have given a relation between the composite 
%representation ($R_c\equiv(R,R)$) invariants ${\cal H}_{(R,R)}(q,\lambda)$
%in $U(N)$ Chern-Simons gauge theory to the $SO(N)$ Chern-Simons invariants 
%${\cal G}_R(q,\lambda)$ of framed knots and links carrying representation $R$. 
\subsection{Composite Invariants in $U(N)$ Chern-Simons Gauge Theory}
The composite representation, $R_c \equiv (R,S)$ labelled by a 
pair of Young diagram is defined as \cite{mar9,compo,mor1,mor2}
\begin{equation}
R_c \equiv (R,S)= {\sum_{U,V,W}} {{(-1)}^{l(U)}}N^R_{U V} N^{S}_{{{U}^T}W} 
(V \times \bar W)~, \label {composite}
\end{equation}
where $U,V,W$ are the representations of the group U(N) , $\ell(U)$ denotes
the number of boxes in the Young diagram corresponding to $U$ and $N$ is
the Littlewood-Richardson coefficient for multiplication of the Young diagrams.

If we take the simplest defining representation 
for $R=\one$ and $S=\one$, then the composite representation $R_c=
(\one,\one)$ derived from eqn. (\ref{composite}) will be the adjoint 
representation of $U(N)$. In terms of fundamental
weights, the highest weight of $R_c$ is $\Lambda^{(1)}+\Lambda^{(N-1)}$. 
Using the above eqn.(\ref {composite}), one can obtain the 
$SU(N)$ representation for any composite representation $(R,S)$ and the 
corresponding highest weight will be $\Lambda_R + \Lambda_{\bar S}$ 
where $\Lambda_R$ and $\Lambda_{\bar S}$
are the highest weights of representation $R$ and conjugate representation
$\bar S$ respectively.

We will now explicitly evaluate the polynomials for various knots
and links carrying the composite representation $(R,S)$ in $U(N)$ Chern-Simons
theory. For the simplest circle called unknot $U_p$ with an arbitrary framing 
$p$, the composite invariant will be framing factor multiplying
the quantum dimension of the composite representation $(R,S)$: 
\begin{equation}
{{\cal H}_{(R,S)}}[U_p] = (-1)^{\ell p} q^{p \{n_{(R,S)}\}^2\over 2}
q^{p C_{(R,S)}}dim_q (R,S)~. \label {frame}
\end{equation}
where $\ell$ is the total number of boxes in the Young diagram
for composite representation $(R,S)$, $C_{(R,S)}$ denotes the $SU(N)$ 
quadratic casimir (\ref {casi})  and 
$n_{(R,S)}$ represents the $U(1)$ charge for the composite representation
$(R,S)$. From the definition of the composite representation highest
weight, it appears that the $U(1)$ charge $n_{(R,S)}$ must be must be
difference of $U(1)$ charges $n_R$ and $n_S$ of representation $R$ and
$S$:
\begin{equation}
n_{(R,S)}=|n_R-n_S|~.\label {framchar}
\end{equation}
The $U(1)$ charges were chosen \cite{mari1,ramprav}
such the $U(N)$ invariants are 
polynomials in two variables $q,\lambda$ \cite{mari1,ramprav}. 
For representation $R$ with $\ell(R)$ number of
boxes in the Young diagram representation, the $U(1)$ charge $n_R$ is
\begin{equation}
n_R={\ell(R) \over \sqrt N}~. \label{u1char}
\end{equation}  
Substituting the  $U(1)$ charge (\ref{u1char}) in eqn.(\ref {framchar}),
the unknot invariant (\ref {frame}) simplies to
\begin{equation}
{{\cal H}_{(R,S)}}[U_p] = (-1)^{\ell p} q^{\kappa_R+\kappa_S} dim_q (R,S)~.
\label {frame1}
\end{equation}
In other words, the framing factor for the knots carrying
composite representation $(R,S)$ involves only the sum of
$\kappa_R$ and $\kappa_S$ as defined in eqn. (\ref {casi}).  
Now, we can write the $U(N)$ framed knot invariants for torus knots
of the type $(2, 2m+1)$ with framing $p$ as follows:
\begin{equation}
{{\cal H}_{(R,S)}}[{\cal K}](q,\lambda)= (-1)^{\ell p} q^{p (\kappa_R+
\kappa_S)} \sum_{R_t}  
dim_q R_t (\lambda_t)^{2m+1}~, \label {knotcom}
\end{equation} 
where $R_t \in (R,S) \otimes (R,S)$ and $\lambda_t$ 
is the braiding eigenvalue in standard framing ($p=0$) for
parallely oriented strands:
\begin{equation}
\lambda_t= \epsilon_t q^{2C_{(R,S)}-C_{R_t}/2}~ ,\label{egval}
\end{equation}
where $\epsilon_{R_t}= \pm 1$ depending upon whether the representation $R_t$
appears symmetrically or antisymmetrically with respect to the tensor
product $(R,S) \otimes (R,S)$ in the $U(N)_k$ Wess-Zumino Witten model.
Unlike the totally symmetric or totally antisymmetric representations,
the tensor product of composite representations does give multiplicities
and we fix the sign of $\epsilon_{R_t}$ by imposing isotopy equivalence of
two or more knots. In fact, fixing the sign of the eigenvalues for
such composite representation was a non-trivial exercise. So,
in appendix A, we have explicitly given all the irreducible representations
$R_t$ and the signs $\epsilon_t$ for some composite representations
so that the composite invariants can be computed. 

The $U(N)$ invariants for framed torus links of the type $(2,2m)$
can also be written. For example, the $U(N)$ invariant for a 
Hopf link of type $(2,2)$ with linking number $-1$ and framing numbers 
$p_1$ and $p_2$ on the component knots carrying representations 
$(R_1,S_1)$ and $(R_2,S_2)$ will be 
\begin{eqnarray}
{\cal H}_{(R_1,S_1),(R_2,S_2)}[H](q,\lambda)&=& (-1)^{\ell_1 p_1 + \ell_2 p_2} 
q^{p_1 (\kappa_{R_1}+\kappa_{S_1})+p_2 (\kappa_{R_2}+\kappa_{S_2})} \times\nonumber\\
~&~&q^{\ell k n_{(R_1,S_1)}n_{(R_2,S_2)}}\sum_{R_t} 
dim_q R_t q^{C_{(R_1,S_1)} + C_{(R_2,S_2)}- C_{R_t}}~,
\end{eqnarray}
where $\ell k=-1$ is the linking number between the two-components
and $R_t \in (R_1,S_1)\otimes(R_2,S_2)$.
We now explicitly evaluate the knot polynomials carrying the composite
representation $(\one,\one)$ in $U(N)$ Chern-Simons theory, for the knots
upto five crossings.
For the simplest composite representation $(\one,\one)$, 
which we denote by  $\rho_0$, the highest weight is 
$\Lambda^{(N-1)}+\Lambda^{(1)}~.$ 
The $p$-frame unknot invariant for this representation is
\begin{equation}
{\cal H}_{(\one,\one)} [U_p]= (-1)^{\ell p}\lambda^{p}(dim_q \rho_0)=
(-1)^{Np}\lambda^{p}[N-1][N+1]~,
\end{equation}
where rewriting the quantum numbers (\ref {qno}) will give the $p$-framed
unknot invariant in variables $q,\lambda=q^N$.
The highest weights for all the representations $R_t$'s obtained from 
$\rho_0 \otimes \rho_0$ and their corresponding
quantum dimensions(\ref{quan}) with the braiding eigenvalues 
(\ref {egval}) are  tabulated below: 

\vskip.1cm
\begin{tabular}{|l|l|l|l|}\hline
$R_t$&Highest weight&Quantum Dimension& $\lambda_t$ \\
\hline
~&~&~&\\ 
$R_1$ & $\Lambda^{(N)}+{\Lambda^{(N-2)}}+2\Lambda^{(1)}$& 
$dim_q R_1 = \frac{[N-1][N-2][N+1][N+2]}{[2][2]}$& $\lambda_1= -\lambda~$\\ 
~&~&~&\\ 
$R_2$& $2\Lambda^{(N-1)}+{\Lambda^{(2)}}$&
$dim_q R_2 = \frac{[N-1][N-2][N+1][N+2]}{[2][2]}$ & $\lambda_2=-\lambda~$\\
~&~&~&\\ 
$R_3$ & $\Lambda^{(N)}+{\Lambda^{(N-2)}}+\Lambda^{(2)}$&
$dim_q R_3 = \frac{[N]^2[N-3][N+1]}{[2][2]}$ & $\lambda_3= q\lambda~$\\
~&~&~&\\ 
$R_4$ & $\Lambda^{(N)}+{\Lambda^{(N-1)}}+\Lambda^{(1)}$&
$dim_q R_4 = [N-1][N+1];$ & $\lambda_4= \lambda^{3/2}$ \\
~&~&~&\\ \hline
\end{tabular}

\begin{tabular}{|l|l|l|l|}\hline
$R_t$&Highest weight&Quantum Dimension& $\lambda_t$ \\
\hline
$R_5$ &$2 \Lambda^{(N)}$&
$dim_q R_5= 1$  & $\lambda_5= \lambda^2~$ \\
~&~&~&\\ 
$R_6$& $2\Lambda^{(N-1)}+2{\Lambda^{(1)}}$&
$dim_q R_6 = \frac{[N]^2[N+3][N-1]}{[2][2]}$ & $\lambda_6= q^{-1} \lambda~$ \\
~&~&~&\\ 
$R_7$ & $\Lambda^{(N)}+{\Lambda^{(N-1)}}+\Lambda^{(1)}$&
$dim_q R_7 = [N-1][N+1]$ & $\lambda_7= -\lambda^{3/2}$\\
\hline 
\end{tabular}
\vskip.2cm
Substituting the tabulated data in eqn.(\ref{knotcom}), the knot 
invariants for the framed trefoil and 5-crossing knots with framing 
$p$ carrying representation $\rho_0 = (\one, \one)$ which
have unknot invariants as an overall factor:
\begin{eqnarray} 
{\cal H}_{(\one,\one)} [{\cal K}_3]&=&{\cal H}_{\one,\one}[U_0]
\left(\lambda^{2}{\lambda^p}(q^{-2}+q^{2}+2)
- \lambda^3(2 q^{-2}+q^{-1}+q-2 q^{2}-2 )\right. \nonumber \\
&~& \left.+ \lambda^4(q^{-2}-2 q^{-1}-2 q+ q^{2}+3 ) + 
\lambda^{5} (q^{-1}+q-2 ) \right)\\
{\cal H}_{(\one,\one)} [{\cal K}_5]&=& {\cal H}_{\one,\one}[U_0]
\left[\lambda^{4}{\lambda^p}(q^{-4}+2q^{-2}+2q^2+q^4+3)\right.\nonumber\\
&&-\lambda^5(2q^{-4}+q^{-3}-4q^{-2}+q^{-1}+q-4q^2+q^3-2q^4-4)\nonumber \\
&& + \lambda^6(q^{-4}-2q^{-3}+3q^{-2}-2q^{-1}-2q+3q^2-2q^3+q^4+4)\nonumber \\
&& +\lambda^7(q^{-3}-2q^{-2}+2q^{-1}+2q-2q^2+q^3-2)\nonumber \\
&&\left. + \lambda^8 (q^{-2}-2q^{-1}-2q+q^2+2)
+ \lambda^9 (q^{-1}+q-2) \right]
\end{eqnarray} 
We have presented the tensor products
and knot invariants for other composite representations in the 
appendix A. These composite invariants play a very crucial role in 
obtaining the topological string amplitudes corresponding to cross-caps
$c=0,1~ {\rm and}~2$. Using these $U(N)$ composite invariants
and the $SO(N)$ invariants in appendix A of Ref.\cite{prav}, we have verified
generalised Rudolph's theorem (\ref {rud}). 
\section{ Reformulated Link Invariants}
We will now review the conjectures proposed by Marino {\cite{mar9}}
for the reformulated $SO(N)$ invariants of knots and links.
Particularly, we have to get the untwisted sector (oriented) contribution
(\ref {orie1})to the open topological string amplitudes on the orientifold
of the resolved conifold geometry. 

Using the properties satisfied by Schur polynomials,
eqn.(\ref {orient}) implies that the oriented invariants
${\cal R}_{R_1,\ldots,R_r}[L]$ of the link $L$ whose
components ${\cal K}_1,\ldots,{\cal K}_r$ are colored by
representations $R_1,\ldots,R_r$ is given by
\begin{equation} 
{\cal R}_{R_1,\ldots,R_r} [L] = \sum_{S_1,T_1,\ldots,S_r,T_r} 
\prod_{\alpha=1}^r N^{R_\alpha}_{S_\alpha,T_\alpha} 
{\cal H}_{(S_1,T_1),\ldots,(S_r,T_r)}[L]~, \label {orie2}
\end{equation}
where $N^{R_\alpha}_{S_\alpha,T_\alpha}$ are the Littlewood-Richardson 
coefficients and ${\cal H}_{(S_1,T_1),\ldots,(S_r,T_r)}[L]$ are composite
invariants in $U(N)$ Chern-Simons gauge theory
of the link whose components carry the composite
representations $(S_1,T_1),\ldots,(S_r,T_r)$ of $U(N)$.
The generating functional giving the oriented contribution to the 
open topological string partition function (\ref {orie1}) is defined as
\begin{equation}
{\cal Z}_{\cal R}(v_1,\ldots,v_r) = \sum_{R_1,\ldots,R_r} 
{\cal R}_{R_1,\ldots,R_r} [L]\,
\prod_{\alpha=1}^r s_{R_{\alpha}}(v_{\alpha}); \,\,
{\cal F}_{\cal R}(v_1,\ldots,v_r) = {\rm log} \, {\cal Z}_{\cal R}(v_1,\ldots,v_r)~,
\end{equation} 
where $s_R(v)$ are the Schur polynomials. Also the generating 
functionals for those involving $SO(N)$ Chern-Simons invariants , 
${\cal G}_{R_1,\ldots,R_r}$, of a link $L$ are defined as
\begin{equation} 
{\cal Z}_{\cal G}(v_1,\ldots,v_r) = \sum_{R_1,\ldots,R_r} 
{\cal G}_{R_1,\ldots,R_r} [L] \, 
\prod_{\alpha=1}^r s_{R_{\alpha}}(v_{\alpha}); \,\,
{\cal F}_{\cal G}(v_1,\ldots,v_r) = {\rm log} \, {\cal Z}_{\cal G}(v_1,\ldots,v_r)~.
\end{equation} 
Marino \cite{mar9} has conjectured a specific form for
these generating functionals:
\begin{equation} 
{\cal F}_{\cal R}(v_1,\ldots,v_r) = \sum_{d=1}^{\infty} \sum_{R_1,\ldots,R_r}
h_{R_1,\ldots,R_r} (q^d,\lambda^d) \, \prod_{\alpha=1}^r s_{R_{\alpha}}
(v_{\alpha}^d)~, \label {con1a}
\end{equation}
and
\begin{equation}
{\cal F}_{\cal G}(v_1,\ldots,v_r) - \frac{1}{2} {\cal F}_{\cal R}(v_1,\ldots,v_r) = \sum_{d\,odd} \sum_{R_1,\ldots,R_r}
g_{R_1,\ldots,R_r} (q^d,\lambda^d) \, \prod_{\alpha=1}^r s_{R_{\alpha}}
(v_{\alpha}^d)~, \label {con2a}
\end{equation}
where $h_{R_1,\ldots,R_r} (q,\lambda)$ and $g_{R_1,\ldots,R_r} (q,\lambda)$
are the reformulated polynomial invariants involving the $U(N)$
and $SO(N)$ Chern-Simons link invariants respectively. The reformulated
invariants are polynomials in $q$ and $\lambda$ and conjectured to 
obey the following form
\begin{equation}
h_{R_1,\ldots,R_r} (q,\lambda) \,{\rm or}\, g_{R_1,\ldots,R_r} (q,\lambda)=
\sum_{Q,s} \frac{1}{q^{1/2}-q^{-1/2}} \tilde{N}_{R_1,\ldots,R_r,Q,s}
q^s\,\lambda^Q\,,\label{horg}
\end{equation}
where $\tilde{N}_{R_1,\ldots,R_r,Q,s}$ are integers.
Though we know that the reformulated invariants
$f_R(q,\lambda)$ obtained from $U(N)$ invariants ${\cal H}_R[L]$ satisfies
the conjecture (\ref {unrfm}), it is not at all obvious that 
the reformulated invariant 
$h_{R_1,\ldots,R_r} (q,\lambda)$ corresponding to 
the oriented invariants (\ref {orie2}) involving linear
combination of $U(N)$ composite invariants must obey a similar
conjectured form (\ref {horg}). We check few examples in section 4 to 
verify Marino's conjecture on the oriented reformulated invariants.
These reformulated invariants are further refined using the following 
equations, in order to reveal the BPS structure
\begin{eqnarray} 
h_{R_1,\ldots,R_r}(q,\lambda) &=& \sum_{S_1,\ldots,S_r}
 M_{R_1,\ldots,R_r;S_1,\ldots,S_r} \, \hat{h}_{S_1,\ldots,S_r}(q,\lambda),\label{reh}\\
g_{R_1,\ldots,R_r}(q,\lambda) &=& \sum_{S_1,\ldots,S_r}
 M_{R_1,\ldots,R_r;S_1,\ldots,S_r} \, \hat{g}_{S_1,\ldots,S_r}(q,\lambda)~,\label{reg}
\end{eqnarray} 
where
\begin{equation}
M_{R_1, \ldots R_r;S_1,\ldots S_r} = \sum_{T_1, \ldots
T_r} \prod_{\alpha=1}^r C_{R_{\alpha} S_{\alpha} T_{\alpha}}
S_{T_{\alpha}}(q)~,
\end{equation}
$R_{\alpha},S_{\alpha}, T_{\alpha}$ are
representations of the symmetric group $S_{\ell_{\alpha}}$ which can be
labelled by a Young-Tableau with a total of $\ell_{\alpha}$ boxes
and $C_{RST}$ are the Clebsch-Gordan coefficients of the symmetric
group. $S_R(q)$ is non-zero only for the hook representations. For a hook
representation having $\ell -d$ boxes in the first row of Young tableau
with total $\ell$ boxes, $S_R(q)=(-1)^d q^{-(\ell-1)/2+d}$. Marino
\cite{mar9} has conjectured that the refined reformulated invariants 
$\hat{h}_{R_1,\ldots,R_r} (q,\lambda)$ 
and $\hat{g}_{R_1,\ldots,R_r} (q,\lambda)$ should have the following structure:
\begin{eqnarray}
\hat{h}_{R_1,\ldots,R_r} (q,\lambda) &=& z^{r-2} \sum_{g\geq 0} \sum_{Q}
N^{c=0}_{R_1,\ldots,R_r,g,Q} z^{2g} \lambda^Q~, \label{hath} \\
\hat{g}_{R_1,\ldots,R_r} (q,\lambda) &=& z^{r-1} \sum_{g\geq 0} \sum_{Q}
\left( N^{c=1}_{R_1,\ldots,R_r,g,Q} z^{2g} \lambda^Q 
+ N^{c=2}_{R_1,\ldots,R_r,g,Q} z^{2g+1} \lambda^Q\right)~,\label{hatg}
\end{eqnarray}
where $N^{c=0}_{R_1,\ldots,R_r,g,Q}$, $N^{c=1}_{R_1,\ldots,R_r,g,Q}$ and
$N^{c=2}_{R_1,\ldots,R_r,g,Q}$ are the BPS invariants corresponding to
cross-caps $c=0,1 \,{\rm and}\, 2$ respectively and the variable
$z=q^{1/2}-q^{-1/2}$.

In the next three sections, we obtain the reformulated invariants and
obtain the BPS integers coefficients for framed knots and framed
two-component links.

\section{Computation of Oriented Invariants $h_{R_1,\ldots,R_r}(q,\lambda)$
and BPS invariants $N^{c=0}_{R_1, \ldots R_r,g,Q}$}
In this section, we list the reformulated oriented invariants and 
the corresponding BPS invariants for simple framed knots 
like unknot and trefoil knot to verify the conjecture (\ref {horg}).

\subsection{Framed unknot}
The reformulated invariants (\ref {horg}) corresponding to the 
oriented invariants for the unknot with framing $p$ are given below
\begin{equation} 
h_{\one}=\frac{1}{(-1 + q)\sqrt{\lambda}} \left( 2\,{\left( -1 \right) }^p\,
\lambda^{p/2}{\sqrt{q}}\,\left( -1 + \lambda  \right) \right)
\end{equation} 
\begin{eqnarray} 
h_{\twohor}&=& \frac{-\lambda^{p-1}}{{\left( -1 + q \right) }^2\,\left( 1 + q \right) } 
\left( \lambda  - 2\,q^{1 + p}\,\left( -1 + \lambda  \right) \,\left( -1 + q\,\lambda  \right)  + 
  {\left( -1 \right) }^p\,\left( -1 + q \right) \,q\,\left( -1 + {\lambda }^2 \right) \right.\nonumber \\
&& \left. + 
  q\,\left( 1 + q + \left( -3 + \left( -3 + q \right) \,q \right) \,\lambda  + 
     \left( 1 + q \right) \,{\lambda }^2 \right) \right)
\end{eqnarray}
\begin{eqnarray} 
h_{\twover}&=&\frac{\lambda^{p-1}}{{\left( -1 + q \right) }^2\,\left( 1 + q \right) }
\left( -2\,q^{1 - p}\,\left( q - \lambda  \right) \,\left( -1 + \lambda  \right)  + 
  {\left( -1 \right) }^p\,\left( -1 + q \right) \,q\,\left( -1 + {\lambda }^2 \right) \right.\nonumber \\
&& \left. - 
  \left( 1 + q \right) \,\left( \lambda  + q\,
      \left( 1 + \lambda \,\left( -4 + q + \lambda  \right)  \right)  \right) \right)
\end{eqnarray}
These results alongwith eqs.(\ref{reh}) and (\ref{hath}) give the
following BPS invariants.\\
\underline{Unknot with framing $p=0$}\\
$$
N^{c=0}_{\one,0,\pm 1/2}=\pm 2, \qquad
N^{c=0}_{\twohor,0,0}=1, \qquad
N^{c=0}_{\twover,0,0}=1.
$$
\underline{Unknot with framing $p=1$}\\
\begin{center}
\begin{tabular}{r|rr} \hline
g & Q=0 & 1 \\ \hline
0 & 2 & -2 \\ \hline
\end{tabular}
\hspace{0.3in}
$N^{c=0}_{\twohor,0,1}=1$
\hspace{0.3in}
\begin{tabular}{r|rr} \hline
g & Q=1 & 2 \\ \hline
0 & 3 & -2 \\ \hline
\end{tabular}

$N^{c=0}_{\one,g,Q}$\hspace{2in}
$N^{c=0}_{\twover,g,Q}$
\end{center}

\underline{Unknot with framing $p=2$}

\begin{center}
\begin{tabular}{r|rr} \hline
g & Q=1/2 & 3/2 \\ \hline
0 & -2 & 2 \\ \hline
\end{tabular}
\hspace{0.5in}
\begin{tabular}{r|rr} \hline
g & Q=2 & 3 \\ \hline
0 & 3 & -2 \\ \hline
\end{tabular}
\hspace{0.5in}
\begin{tabular}{r|rrr} \hline
g & Q=1 & 2 & 3 \\ \hline
0 & -2 & 7 & -4 \\
1 & 0 & 2 & -2 \\ \hline
\end{tabular}

$N^{c=0}_{\one,g,Q}$ \hspace{1in}
$N^{c=0}_{\twohor,g,Q}$ \hspace{1in}
$N^{c=0}_{\twover,g,Q}$
\end{center}

\subsection{Framed trefoil knot}

For the trefoil knot with framing $p$, the oriented invariants are
\begin{eqnarray}
h_{\one}=\frac{1}{\left( -1 + q \right) \,{\sqrt{q}}}
\left( 2\,{\left( -1 \right) }^p\,\lambda^{p/2}\left( -1 + \lambda  \right) \,{\sqrt{\lambda }}\,\left( 1 + q^2 - q\,\lambda  \right) \right)
\end{eqnarray} 
\begin{eqnarray} 
h_{\twohor}&=&\frac{\lambda^{p+1}}{{\left( -1 + q \right) }^2\,q^2\,\left( 1 + q \right) }
\left( -2\,q^{1 + p}\,\left( -1 + \lambda  \right)  + 2\,q^{2 + p}\,\left( -1 + \lambda  \right) \,\lambda  + 
    2\,q^{3 + p}\,\left( -1 + \lambda  \right) \,\lambda \right.\nonumber \\
&& - {\left( -1 + \lambda  \right) }^2\,\lambda  - 
    q^7\,{\left( -1 + \lambda  \right) }^2\,\lambda  - 2\,q^{8 + p}\,{\left( -1 + \lambda  \right) }^2\,\lambda  + 
    q\,\left( 1 + {\left( -1 \right) }^p - \lambda  \right) \,\left( -1 + {\lambda }^2 \right) \nonumber \\
&& - 
    q^6\,\left( -1 + {\left( -1 \right) }^p + \lambda  \right) \left( -1 + {\lambda }^2 \right)  + 
    2\,q^{7 + p}\,{\left( -1 + \lambda  \right) }^2\left( 1 + {\lambda }^2 \right)  - 
    2\,q^{6 + p}\,\lambda \,\left( 2 - 3\,\lambda  + {\lambda }^2 \right) \nonumber \\
&& - 
    2\,q^{4 + p}\,\left( -1 + 2\,\lambda  - 2\,{\lambda }^2 + {\lambda }^3 \right)  - 
    2\,q^{5 + p}\,\left( -1 + 2\,\lambda  - 2\,{\lambda }^2 + {\lambda }^3 \right) \nonumber \\
&& + 
    q^5\,\left( -1 - {\left( -1 \right) }^p + 4\,\lambda  + \left( -7 + {\left( -1 \right) }^p \right) \,{\lambda }^2 + 
       2\,{\lambda }^3 + {\lambda }^5 \right) \nonumber \\
&& + 
    q^2\,\left( -1 + {\left( -1 \right) }^p + 4\,\lambda  - \left( 7 + {\left( -1 \right) }^p \right) \,{\lambda }^2 + 
       2\,{\lambda }^3 + {\lambda }^5 \right) \nonumber \\
&& - 
    q^4\,\left( 2 - 8\,\lambda  + \left( 9 + {\left( -1 \right) }^p \right) \,{\lambda }^2 - 6\,{\lambda }^3 - 
       \left( -1 + {\left( -1 \right) }^p \right) \,{\lambda }^4 + {\lambda }^5 \right) \nonumber \\
&& \left. - 
    q^3\,\left( 2 - 8\,\lambda  - \left( -9 + {\left( -1 \right) }^p \right) \,{\lambda }^2 - 6\,{\lambda }^3 + 
       \left( 1 + {\left( -1 \right) }^p \right) \,{\lambda }^4 + {\lambda }^5 \right)  \right) 
\end{eqnarray} 
\begin{eqnarray} 
h_{\twover}&=&\frac{\lambda^{p+1}}{{\left( -1 + q \right) }^2\,q^3\,\left( 1 + q \right) }
\left( -2\,q^{7-p}\,\left( -1 + \lambda  \right)  + 2\,q^{5-p}\,\left( -1 + \lambda  \right) \,\lambda  + 
      2\,q^{6-p}\,\left( -1 + \lambda  \right) \,\lambda\right.\nonumber \\
&&  - 2\,q^{-p}{\left( -1 + \lambda  \right) }^2\,\lambda  - 
      q\,{\left( -1 + \lambda  \right) }^2\,\lambda  - q^8\,{\left( -1 + \lambda  \right) }^2\,\lambda  + 
      q^7\,\left( 1 + {\left( -1 \right) }^p - \lambda  \right) \,\left( -1 + {\lambda }^2 \right) \nonumber \\
&& - 
      q^2\,\left( -1 + {\left( -1 \right) }^p + \lambda  \right) \,\left( -1 + {\lambda }^2 \right)  + 
      2\,q^{1-p}\,{\left( -1 + \lambda  \right) }^2\,\left( 1 + {\lambda }^2 \right)  - 
      2\,q^{2-p}\,\lambda \,\left( 2 - 3\,\lambda  + {\lambda }^2 \right) \nonumber \\
&& - 
      2\,q^{3-p}\,\left( -1 + 2\,\lambda  - 2\,{\lambda }^2 + {\lambda }^3 \right)  - 
      2\,q^{4-p}\,\left( -1 + 2\,\lambda  - 2\,{\lambda }^2 + {\lambda }^3 \right) \nonumber \\
&& + 
      q^3\,\left( -1 - {\left( -1 \right) }^p + 4\,\lambda  + 
         \left( -7 + {\left( -1 \right) }^p \right) \,{\lambda }^2 + 2\,{\lambda }^3 + {\lambda }^5 \right) \nonumber \\
&& + 
      q^6\,\left( -1 + {\left( -1 \right) }^p + 4\,\lambda  - 
         \left( 7 + {\left( -1 \right) }^p \right) \,{\lambda }^2 + 2\,{\lambda }^3 + {\lambda }^5 \right) \nonumber \\
&& - 
      q^4\,\left( 2 - 8\,\lambda  + \left( 9 + {\left( -1 \right) }^p \right) \,{\lambda }^2 - 6\,{\lambda }^3 - 
         \left( -1 + {\left( -1 \right) }^p \right) \,{\lambda }^4 + {\lambda }^5 \right) \nonumber \\
&& \left. - 
      q^5\,\left( 2 - 8\,\lambda  - \left( -9 + {\left( -1 \right) }^p \right) \,{\lambda }^2 - 6\,{\lambda }^3 + 
         \left( 1 + {\left( -1 \right) }^p \right) \,{\lambda }^4 + {\lambda }^5 \right)  \right)
\end{eqnarray}
From eqns.(\ref{reh}) and (\ref{hath}) we obtain the BPS invariants
corresponding to cross-cap $c=0$.\\
\underline{Trefoil knot with framing $p=0$}

\begin{center} 
\begin{tabular}{r|rrr} \hline
g & Q=1/2 & 3/2 & 5/2 \\ \hline
0 & -4 & 6 & -2 \\
1 & -2 & 2 & 0 \\ \hline
\end{tabular}

$N^{c=0}_{\one,g,Q}$
\end{center}  

\begin{center} 
\begin{tabular}{r|rrrrrr} \hline
g & Q=1 & 2 & 3 & 4 & 5 & 6 \\ \hline
0 & -4 & 25 & -40 & 25 & -4 & -1 \\
1 & -2 & 18 & -32 & 18 & -2 & 0 \\
2 & 0 & 3 & -6 & 3 & 0 & 0 \\ \hline
\end{tabular}
\hspace{0.5in}
\begin{tabular}{r|rrrrrr} \hline
g & Q=1 & 2 & 3 & 4 & 5 & 6 \\ \hline
0 & -8 & 41 & -64 & 41 & -8 & -1 \\
1 & -8 & 46 & -76 & 46 & -8 & 0 \\
2 & -2 & 17 & -30 & 17 & -2 & 0 \\
3 & 0 & 2 & -4 & 2 & 0 & 0 \\ \hline
\end{tabular}

$N^{c=0}_{\twohor,g,Q}$ \hspace{1.5in}
$N^{c=0}_{\twover,g,Q}$
\end{center}  

\underline{Trefoil knot with framing $p=1$}

\begin{center} 
\begin{tabular}{r|rrr} \hline
g & Q=1 & 2 & 3 \\ \hline
0 & 4 & -6 & 2 \\
1 & 2 & -2 & 0 \\ \hline
\end{tabular}

$N^{c=0}_{\one,g,Q}$
\end{center}  

\begin{center} 
\begin{tabular}{r|rrrrrr} \hline
g & Q= 2 & 3 & 4 & 5 & 6 & 7 \\ \hline
0 & -8 & 41 & -64 & 41 & -8 & -1 \\
1 & -8 & 46 & -76 & 46 & -8 & 0 \\
2 & -2 & 17 & -30 & 17 & -2 & 0 \\
3 & 0 & 2 & -4 & 2 & 0 & 0 \\ \hline
\end{tabular}
\hspace{0.5in}
\begin{tabular}{r|rrrrrr} \hline
g & Q=2 & 3 & 4 & 5 & 6 & 7\\ \hline
0 & -14 & 69 & -104 & 63 & -12 & -1 \\
1 & -22 & 116 & -180 & 108 & -22 & 0 \\
2 & -12 & 73 & -120 & 71 & -12 & 0 \\
3 & -2 & 20 & -36 & 20 & -2 & 0 \\
4 & 0 & 2 & -4 & 2 & 0 & 0 \\ \hline
\end{tabular}

$N^{c=0}_{\twohor,g,Q}$ \hspace{2.5in}
$N^{c=0}_{\twover,g,Q}$
\end{center}

\underline{Trefoil knot with framing $p=2$}

\begin{center} 
\begin{tabular}{r|rrr} \hline
g & Q=3/2 & 5/2 & 7/2 \\ \hline
0 & -4 & 6 & -2 \\
1 & -2 & 2 & 0 \\ \hline
\end{tabular}

$N^{c=0}_{\one,g,Q}$
\end{center}

\begin{center} 
\begin{tabular}{r|rrrrrr} \hline
g & Q=3 & 4 & 5 & 6 & 7 & 8 \\ \hline
0 & -14 & 69 & -104 & 63 & -12 & -1 \\
1 & -22 & 116 & -180 & 108 & -22 & 0 \\
2 & -12 & 73 & -120 & 71 & -12 & 0 \\
3 & -2 & 20 & -36 & 20 & -2 & 0 \\
4 & 0 & 2 & -4 & 2 & 0 & 0 \\ \hline
\end{tabular}
\hspace{0.2in}
\begin{tabular}{r|rrrrrr} \hline
g & Q=3 & 4 & 5 & 6 & 7 & 8 \\ \hline
0 & -26 & 109 & -154 & 91 & -18 & -1 \\
1 & -58 & 256 & -376 & 226 & -48 & 0 \\
2 & -46 & 241 & -378 & 227 & -44 & 0 \\
3 & -16 & 110 & -186 & 108 & -16 & 0 \\
4 & -2 & 24 & -44 & 24 & -2 & 0 \\
5 & 0 & 2 & -4 & 2 & 0 & 0 \\ \hline
\end{tabular}

$N^{c=0}_{\twohor,g,Q}$ \hspace{2.5in}
$N^{c=0}_{\twover,g,Q}$
\end{center} 

\section{Explicit Computation of $SO(N)$ reformulated invariants 
$g_{R_1,R_2, \ldots R_r}(q,\lambda)$}

In this section we compute the functions $g_{R_1, \ldots
R_r}(q,\lambda)$ for various nontrivial framed knots and links and
show that they obey the conjectured  
form (\ref{horg}).
\subsection{Framed Unknot}

\begin{equation}
g_{\one}={\left( -1 \right) }^p\,\lambda^{p/2}
\end{equation}

\begin{equation}
g_{\twohor}=\frac{1}{\left( -1 + q \right)} \lambda^{p-1/2}\, \left( -{\sqrt{q}} + q^{\frac{1}{2} + p} \right) \,\left( -1 + \lambda  \right)
\end{equation}

\begin{equation}
g_{\twover}=- \frac{q^{-p}\,\lambda^{p-1/2}}{\left( -1 + q \right)} \left( -{\sqrt{q}} + q^{\frac{1}{2} + p} \right) \,\left( -1 + \lambda  \right)
\end{equation}

\begin{eqnarray}
g_{\threehor} &=& \frac{\lambda^{3p/2-1}}{{\left( -1 + q \right) }^2\,\left( 1 + q \right) }
\left( {\left( -1 \right) }^p\,q\,\left( -1 + q^p \right) \,\left( -1 + \lambda  \right) \right.\nonumber \\
&& \left.  \left( 1 + q - q^p - q^{2\,p}  - \lambda  - q\,\lambda  + q^{1 + p}\,\lambda  + q^{1 + 2\,p}\,\lambda  \right) \right)
\end{eqnarray}

\begin{eqnarray}
g_{\mixed} &=& \frac{-\lambda^{3p/2-1}}{{\left( -1 + q \right) }^2\,\left( 1 + q \right) }
{\left( -1 \right) }^p\,q^{1 - p}\,\left( -1 + q^p \right) \,\left( -1 + \lambda  \right) \,
  \left( 1 - q\,\left( -2 + \lambda  \right)\right.\nonumber \\
&& \left.  + q^p\,\left( -2 + \lambda  \right)  - 2\,\lambda  + 
    q^{1 + p}\,\left( -1 + 2\,\lambda  \right)  \right)
\end{eqnarray}

\begin{eqnarray}
g_{\threever} &=& \frac{\lambda^{3p/2-1}}{{\left( -1 + q \right) }^2\,\left( 1 + q \right) }
{\left( -1 \right) }^p\,q^{\frac{1}{2} - 3\,p}\,\left( -1 + \lambda  \right) \,
  \left( -q^{\frac{3}{2}} + q^{\frac{1}{2} + 2\,p}\,\left( 1 - 2\,\lambda  \right) \right.\nonumber \\
&& \left. - 
    q^{\frac{3}{2} + 2\,p}\,\left( -2 + \lambda  \right)  + 
    q^{\frac{1}{2} + 3\,p}\,\left( -1 + \lambda  \right)  + 
    q^{\frac{3}{2} + 3\,p}\,\left( -1 + \lambda  \right)  + {\sqrt{q}}\,\lambda  \right)
\end{eqnarray}
\subsection{Framed Trefoil Knot}

\begin{equation}
g_{\one}={\left( -1 \right) }^p\,q^{-1}\,\lambda^{p/2+1} \,\left( 1 - q^2\,\left( -1 + \lambda  \right)  - \lambda  + 
      q\,\left( 1 - \lambda  + {\lambda }^2 \right)  \right) 
\end{equation}

\begin{eqnarray}
g_{\twohor} &=& \frac{\lambda^{p}}{\left( -1 + q \right) \,q^2} \left( -1 + \lambda  \right) \,{\lambda }^{\frac{3}{2}}\,
    \left( q^{\frac{1}{2} + p} + q^{\frac{9}{2} + p}\,\left( 1 - 2\,\lambda  \right)  + 
      {\sqrt{q}}\,\left( -1 + \lambda  \right) \right.\nonumber \\
&& + q^{\frac{9}{2}}\,\left( -1 + \lambda  \right)  - 
      q^{\frac{5}{2} + p}\,\left( -1 + \lambda  \right)  - q^{\frac{3}{2} + p}\,\lambda  + 
      q^{\frac{15}{2} + p}\,\left( -1 + \lambda  \right) \,\lambda\nonumber \\
&& + q^{\frac{19}{2} + p}\,{\lambda }^2 - 
      q\,{\lambda }^{\frac{5}{2}} + q^4\,{\lambda }^{\frac{5}{2}} + q^{7 + p}\,{\lambda }^{\frac{5}{2}} - 
      q^{10 + p}\,{\lambda }^{\frac{5}{2}} + q^{\frac{3}{2}}\,\left( -1 + 2\,\lambda  - 2\,{\lambda }^2 \right) \nonumber \\
&& + 
      q^{\frac{7}{2}}\,\left( -1 + 2\,\lambda  - 2\,{\lambda }^2 \right)  - 
      q^{\frac{17}{2} + p}\,\lambda \,\left( 1 + {\lambda }^2 \right) + 
      q^{\frac{7}{2} + p}\,\left( 1 - \lambda  + {\lambda }^2 \right) \nonumber \\ 
&& + 
      q^{\frac{11}{2} + p}\,\left( 1 - \lambda  + {\lambda }^2 \right)  + 
      q^{\frac{13}{2} + p}\,\left( 1 - \lambda  + {\lambda }^2 \right)  + 
      q^2\,{\lambda }^{\frac{3}{2}}\,\left( 1 + \lambda  + {\lambda }^2 \right)\nonumber \\
&&  - 
      q^3\,{\lambda }^{\frac{3}{2}}\,\left( 1 + \lambda  + {\lambda }^2 \right)  - 
      q^{8 + p}\,{\lambda }^{\frac{3}{2}}\,\left( 1 + \lambda  + {\lambda }^2 \right)  + 
      q^{9 + p}\,{\lambda }^{\frac{3}{2}}\,\left( 1 + \lambda  + {\lambda }^2 \right) \nonumber \\
&& \left. + q^{\frac{5}{2}}\,\left( -2 + 3\,\lambda  - {\lambda }^2 + {\lambda }^3 \right)  \right) 
\end{eqnarray}

\begin{eqnarray}
g_{\twover} &=& \frac{\lambda^{p}}{-1 + q}\, q^{-7 - p}\,\left( -1 + \lambda  \right) \,{\lambda }^{\frac{3}{2}}\,
    \left( q^{\frac{19}{2}} + q^{\frac{11}{2}}\,\left( 1 - 2\,\lambda  \right)  - 
      q^{\frac{15}{2}}\,\left( -1 + \lambda  \right) \right. \nonumber \\
&&  + q^{\frac{11}{2} + p}\,\left( -1 + \lambda  \right)  + 
      q^{\frac{19}{2} + p}\,\left( -1 + \lambda  \right)  - q^{\frac{17}{2}}\,\lambda  + 
      q^{\frac{5}{2}}\,\left( -1 + \lambda  \right) \,\lambda  + {\sqrt{q}}\,{\lambda }^2 \nonumber \\
&& + {\lambda }^{\frac{5}{2}} - 
      q^3\,{\lambda }^{\frac{5}{2}} - q^{6 + p}\,{\lambda }^{\frac{5}{2}} + q^{9 + p}\,{\lambda }^{\frac{5}{2}} + 
      q^{\frac{13}{2} + p}\,\left( -1 + 2\,\lambda  - 2\,{\lambda }^2 \right)\nonumber \\
&&  + 
      q^{\frac{17}{2} + p}\,\left( -1 + 2\,\lambda  - 2\,{\lambda }^2 \right)  + 
      q^{\frac{7}{2}}\,\left( 1 - \lambda  + {\lambda }^2 \right)  + 
      q^{\frac{9}{2}}\,\left( 1 - \lambda  + {\lambda }^2 \right) \nonumber \\
&& + 
      q^{\frac{13}{2}}\,\left( 1 - \lambda  + {\lambda }^2 \right)  - 
      q\,{\lambda }^{\frac{3}{2}}\,\left( 1 + \lambda  + {\lambda }^2 \right)  + 
      q^2\,{\lambda }^{\frac{3}{2}}\,\left( 1 + \lambda  + {\lambda }^2 \right)  \nonumber \\
&& + 
      q^{7 + p}\,{\lambda }^{\frac{3}{2}}\,\left( 1 + \lambda  + {\lambda }^2 \right)  - 
      q^{8 + p}\,{\lambda }^{\frac{3}{2}}\,\left( 1 + \lambda  + {\lambda }^2 \right)  - 
      q^{\frac{3}{2}}\,\left( \lambda  + {\lambda }^3 \right) \nonumber \\
&& \left. + 
      q^{\frac{15}{2} + p}\,\left( -2 + 3\,\lambda  - {\lambda }^2 + {\lambda }^3 \right)  \right)
\end{eqnarray}

Substituting values for $p$, the above equations reduce to the conjectured
form (\ref {horg}).
We have presented the reformulated invariants for 
few framed knots and two component links in appendix B.

\section{$N_{(R_1, \ldots R_r),g,Q}^{c=1}$ and $N_{(R_1, \ldots R_r),g,Q}^{c=2}$ Computation}

We shall now  compute the integer coefficients corresponding
to cross-cap $c=1$ and $c=2$ unoriented open string amplitude obtained
from $SO(N)$ reformulated invariants for various framed knots and framed
links using eqns.(\ref {reg}, \ref {hatg}).
\subsection{Framed Knots}

\underline{For unknot with zero framing},
the only non zero coefficient is $N_{\one,0,0}^{c=1}=1$.\\ \\
\underline{Unknot with framing $p=1$}
\begin{equation}
N_{\one,0,1/2}^{c=1}=-1\,.
\end{equation}
\begin{center}
\begin{tabular}{r|rr} \hline
g & Q=1/2 & 3/2 \\ \hline
0 & 1 & -1 \\ \hline
\end{tabular}

$N^{c=1}_{\twover,g,Q}$ for the unknot
\end{center}
\underline{Unknot with framing $p=2$}
\begin{equation}
N^{c=1}_{\one,0,1}=1\,.
\end{equation}
\begin{center} 
\begin{tabular}{r|rr} \hline
g & Q=3/2 & 5/2 \\ \hline
0 & 1 & -1 \\ \hline
\end{tabular}
\hspace{1in}
\begin{tabular}{r|rr} \hline
g & Q=3/2 & 5/2 \\ \hline
0 & 3 & -3 \\
1 & 1 & -1 \\ \hline
\end{tabular}

$N^{c=1}_{\twohor,g,Q}$ for the unknot\hspace{0.5in}
$N^{c=1}_{\twover,g,Q}$ for the unknot
\end{center}  

\underline{Trefoil knot with framing $p=1$}

\begin{center}
\begin{tabular}{r|rrrrr} \hline
g & Q=5/2 & 7/2 & 9/2 & 11/2 & 13/2 \\ \hline
0 & 16 & -69 & 111 &  -79 & 21 \\
1 & 20 & -146 & 307 & -251 & 70 \\
2 & 8 & -128 & 366 & -330 & 84 \\
3 & 1 & -56 & 230 & -220 & 45 \\
4 & 0 & -12 & 79 & -78 & 11 \\
5 & 0 & -1 &  14 & -14 & 1 \\
6 & 0 & 0 & 1 & -1 & 0 \\ \hline
\end{tabular}
\hspace{.1in}
\begin{tabular}{r|rrrr} \hline
g & Q= 4 & 5 & 6 & 7 \\ \hline
0 & 21 & -63 & 63 & -21 \\
1 & 70 & -231 & 231 & -70 \\
2 & 84 & -322 & 322 & -84 \\
3 & 45 & -219 & 219 & -45 \\
4 & 11 & -78 & 79 & -11 \\
5 & 1 &-14 & 14 & -1 \\
6 & 0 & -1 & 1 & 0 \\ \hline
\end{tabular}

$N^{c=1}_{\twohor,g,Q}$ for the trefoil knot \hspace{1in}
$N^{c=2}_{\twohor,g,Q}$ for the trefoil knot
\end{center}

\begin{center}
\begin{tabular}{r|rrrrr} \hline
g & Q= 5/2 & 7/2 & 9/2 & 11/2 & 13/2 \\ \hline
0 & 30 & -114 & 167 & -111 & 28 \\
1 & 55 & -311 & 587 & -457 & 126 \\
2 & 36 & -367 & 912 & -791 & 210 \\
3 & 10 & -230 & 770 & -715 & 165 \\
4 & 1 & -79 & 376 & -364 & 66 \\
5 & 0 & -14 & 106 & -105 & 13 \\
6 & 0 & -1 & 16 & -16 & 1 \\
7 & 0 & 0 & 1 & -1 & 0 \\ \hline
\end{tabular}
\hspace{.1in}
\begin{tabular}{r|rrrr} \hline
g & Q= 4 & 5 & 6 & 7 \\ \hline
0 & 28 & -84 & 84 & -28 \\
1 & 126 & -406 & 406 & -126 \\
2 & 210 & -756 & 756 & -210 \\
3 & 165 & -705 & 705 & -165 \\
4 & 66 & -363 & 363 & -66 \\
5 & 13 & -105 & 105 & -13 \\
6 & 1 & -16 & 16 & -1 \\
7 & 0 & -1 & 1 & 0 \\ \hline
\end{tabular}

$N^{c=1}_{\twover,g,Q}$ for the trefoil knot \hspace{1in}
$N^{c=2}_{\twover,g,Q}$ for the trefoil knot
\end{center}

\underline{Trefoil knot with framing $p=2$}

\begin{center}
\begin{tabular}{r|rrrrr} \hline
g & Q= 7/2 & 9/2 & 11/2 & 13/2 & 15/2 \\ \hline
0 & 30 & -114 & 167 & -111 & 28 \\
1 & 55 & -311 & 587 & -457 & 126 \\
2 & 36 & -367 & 912 & -791 & 210 \\
3 & 10 & -230 & 770 & -715 & 165 \\
4 & 1 & -79 & 376 & -364 & 66 \\
5 & 0 & -14 & 106 & -105 & 13 \\
6 & 0 & -1 & 16 & -16 & 1 \\
7 & 0 & 0 & 1 & -1 & 0 \\ \hline
\end{tabular}
\hspace{.1in}
\begin{tabular}{r|rrrr} \hline
g & Q=5 & 6 & 7 & 8 \\ \hline
0 & 28 & -84 & 84 & -28 \\
1 & 126 & -406 & 406 & -126 \\
2 & 210 & -756 & 756 & -210 \\
3 & 165 & -705 & 705 & -165 \\
4 & 66 & -363 & 363 & -66 \\
5 & 13 & -105 & 105 & -13 \\
6 & 1 & -16 & 16 & -1 \\
7 & 0 & -1 & 1 & 0 \\ \hline
\end{tabular}

$N^{c=1}_{\twohor,g,Q}$ for the trefoil knot \hspace{1in}
$N^{c=2}_{\twohor,g,Q}$ for the trefoil knot
\end{center}

\begin{center}
\begin{tabular}{r|rrrrr} \hline
g & Q= 7/2 & 9/2 & 11/2 & 13/2 & 15/2 \\ \hline
0 & 50 & -174 & 237 & -149 & 36 \\
1 & 125 & -601 & 1042 & -776 & 210 \\
2 & 120 & -919 & 2046 & -1709 & 462 \\
3 & 55 & -771 & 2222 & -2001 & 495 \\
4 & 12 & -376 & 1443 & -1365 & 286 \\
5 & 1 & -106 & 574 & -560 & 91 \\
6 & 0 & -16 & 137 & -136 & 15 \\
7 & 0 & -1 & 18 & -18 & 1 \\
8 & 0 & 0 & 1 & -1 & 0 \\ \hline
\end{tabular}
\hspace{.1in}
\begin{tabular}{r|rrrr} \hline
g & Q= 5 & 6 & 7 & 8 \\ \hline
0 & 36 & -108 & 108 & -36 \\
1 & 210 & -666 & 666 & -210 \\
2 & 462 & -1596 & 1596 & -462 \\
3 & 495 & -1947 & 1947 & -495 \\
4 & 286 & -1353 & 1353 & -286 \\
5 & 91 & -559 & 559 & -91 \\
6 & 15 & -136 & 136 & -15 \\
7 & 1 & -18 & 18 & -1 \\
8 & 0 & -1 & 1 & 0 \\ \hline
\end{tabular}

$N^{c=1}_{\twover,g,Q}$ for the trefoil knot \hspace{1in}
$N^{c=2}_{\twover,g,Q}$ for the trefoil knot
\end{center}

\underline{Torus knot $(2,5)$ with framing $p=0$} 

\begin{center}
\begin{tabular}{r|rrr} \hline
g & Q=1 & 2 & 3 \\ \hline
0 & 3 & -3 & 1 \\
1 & 1 & -1 & 0 \\ \hline
\end{tabular}

$N^{c=1}_{\one,g,Q}$
\end{center}

\begin{center}
\begin{tabular}{c|rrrrrr} \hline
g & $Q$=7/2 & 9/2 & 11/2 & 13/2 & 15/2 & 17/2 \\ \hline
0 & 80 & -285 & 285 & 55 & -225 & 90 \\
1 & 260 & -1190 & 1190 & 910 & -1875 & 705 \\
2 & 336 & -2192 & 2192 & 3801 & -6315 & 2178 \\
3 & 221 & -2286 & 2286 & 7666 & -11385 & 3498 \\
4 & 78 & -1456 & 1456 & 8997& -12364 & 3289 \\
5 & 14 & -575 & 575 & 6642 & -8567 & 1911 \\
6 & 1 & -137 & 137 & 3180 & -3876 & 695 \\
7 & 0 & -18 & 18 & 986 & -1140 & 154 \\
8 & 0 & -1 & 1 & 191 & -210 & 19 \\
9 & 0 & 0 & 0 & 21 & -22 & 1 \\
10 & 0 & 0 & 0 & 1 & -1 & 0 \\ \hline
\end{tabular}

$N^{c=1}_{\twohor,g,Q}$ for torus knot $(2,5)$
\end{center} 

\begin{center} 
\begin{tabular}{c|rrrr} \hline
g & Q=5 & 7 & 8 & 10 \\ \hline
0 & 45 & -225 & 225 & -45 \\
1 & 330 & -1875 & 1875 & -330 \\
2 & 924 & -6315 & 6315 & -924 \\
3 & 1287 & -11385 & 11385 & -1287 \\
4 & 1001 & -12364 & 12364 & -1001 \\
5 & 455 & -8567 & 8567 & -455 \\
6 & 120 & -3876 & 3876 & -120 \\
7 & 17 & -1140 & 1140 & -17 \\
8 & 1 & -210 & 210 & -1 \\
9 & 0 & -22 & 22 & 0 \\
10 & 0 & -1 & 1 & 0 \\ \hline
\end{tabular}

$N^{c=2}_{\twohor,g,Q}$ for torus knot $(2,5)$
\end{center}  
These results are agreeing with the results in Ref.\cite {stev}.

\begin{center} 
\begin{tabular}{r|rrrrrr} \hline
g & Q=7/2 & 9/2 & 11/2 & 13/2 & 15/2 & 17/2 \\ \hline
0 & 120 & -415 & 415 & 45 & -275 & 110 \\
1 & 490 & -2085 & 2085 & 1215 & 2750 & 1045 \\
2 & 819 & -4663 & 4663 & 6364 & -11110 & 3927 \\
3 & 724 & -5994 & 5994 & 15644 & -24090 & 7722 \\
4 & 365 & -4822 & 4822 & 22372 & -31746 & 9009 \\
5 & 105 & -2500 & 2500 & 20370 & -27118 & 6643 \\
6 & 16 & -833 & 833 & 12307 & -15503 & 3180 \\
7 & 1 & -172 & 172 & 4998 & -5985 & 986 \\
8 & 0 & -20 & 20 & 1349 & -1540 & 191 \\
9 & 0 & -1 & 1 & 232 & -253 & 21 \\
10 & 0 & 0 & 0 & 23 & -24 & 1 \\
11 & 0 & 0 & 0 & 1 & -1 & 0 \\ \hline
\end{tabular}

$N^{c=1}_{\twover,g,Q}$ for torus knot $(2,5)$
\end{center}

\begin{center} 
\begin{tabular}{r|rrrr} \hline
g & Q=5 & 7 & 8 & 10 \\ \hline
0 & 55 & -275 & 275 & -55 \\
1 & 495 & -2750 & 2750 & -495 \\
2 & 1716 & -11110 & 11110 & -1716 \\
3 & 3003 & -24090 & 24090 & -3003 \\
4 & 3003 & -31746 & 31746 & -3003 \\
5 & 1820 & -27118 & 27118 & -1820 \\
6 & 680 & -15503 & 15503 & -680 \\
7 & 153 & -5985 & 5985 & -153 \\
8 & 19 & -1540 & 1540 & -19 \\
9 & 1 & -253 & 253 & -1 \\
10 & 0 & -24 & 24 & 0 \\
11 & 0 & -1 & 1 & 0 \\ \hline
\end{tabular}

$N^{c=2}_{\twover,g,Q}$ for torus knot $(2,5)$
\end{center}  

\underline{Torus knot $(2,5)$ with framing $p=1$}

\begin{center}
\begin{tabular}{r|rrr} \hline
g & Q=3/2 & 5/2 & 7/2 \\ \hline
0 & -3 & 3 & -1 \\
1 & -1 & 1 & 0 \\ \hline
\end{tabular}

$N^{c=1}_{\one,g,Q}$
\end{center} 

\begin{center} 
\begin{tabular}{r|rrrrrr} \hline
g & Q= 9/2 & 11/2 & 13/2 & 15/2 & 17/2 & 19/2 \\ \hline
0 & 120 & -415 & 415 & 45 & -275 & 110\\
1 & 490 & -2085 & 2085 & 1215 & -2750 & 1045 \\
2 & 819 & -4663 & 4663 & 6364 & -11110 & 3927 \\
3 & 724 & -5994 & 5994 & 15644 & -24090 & 7722 \\
4 & 365 & -4822 & 4822 & 22372 & -31746 & 9009 \\
5 & 105 & -2500 & 2500 & 20370 & -27118 & 6643 \\
6 & 16 & -833 & 833 & 12307 & -15503 & 3180 \\
7 & 1 & -172 & 172 & 4998 & -5985 & 986 \\
8 & 0 & -20 & 20 & 1349 & -1540 & 191 \\
9 & 0 & -1 & 1 & 232 & -253 & 21 \\
10 & 0 & 0 & 0 & 23 & -24 & 1 \\
11 & 0 & 0 & 0 & 1 & -1 & 0 \\ \hline
\end{tabular}

$N^{c=1}_{\twohor,g,Q}$ for torus knot $(2,5)$
\end{center}  

\begin{center} 
\begin{tabular}{r|rrrr} \hline
g & Q=6 & 8 & 9 & 11 \\ \hline
0 & 55 & -275 & 275 & -55 \\
1 & 495 & -2750 & 2750 & -495 \\
2 & 1716 & -11110 & 11110 & -1716 \\
3 & 3003 & -24090 & 24090 & -3003 \\
4 & 3003 & -31746 & 31746 & -3003 \\
5 & 1820 & -27118 & 27118 & -1820 \\
6 & 680 & -15503 & 15503 & -680 \\
7 & 153 & -5985 & 5985 & -153 \\
8 & 19 & -1540 & 1540 & -19 \\
9 & 1 & -253 & 253 & -1 \\
10 & 0 & -24 & 24 & 0 \\
11 & 0 & -1 & 1 & 0 \\ \hline
\end{tabular}

$N^{c=2}_{\twohor,g,Q}$ for torus knot $(2,5)$
\end{center}  

\begin{center} 
\begin{tabular}{r|rrrrrr} \hline
g & Q= 9/2 & 11/2 & 13/2 & 15/2 & 17/2 & 19/2 \\ \hline
0 & 175 & -585 & 580 & 28 & -330 & 132 \\
1 & 875 & -3480 & 3460 & 1554 & -3905 & 1496 \\
2 & 1820 & -9282 & 9261 & 10136 & -18656 & 6721 \\
3 & 2055 & -14384 & 14376 & 29985 & -47905 & 15873 \\
4 & 1377 & -14184 & 14183 & 51391 & -75218 & 22451 \\
5 & 561 & -9247 & 9247 & 56470 & -77415 & 20384 \\
6 & 136 & -4029 & 4029 & 41804 & -54248 & 12308 \\
7 & 18 & -1159 & 1159 & 21317 & -26333 & 4998 \\
8 & 1 & -211 & 211 & 7505 & -8855 & 1349 \\
9 & 0 & -22 & 22 & 1792 & -2024 & 232 \\
10 & 0 & -1 & 1 & 277 & -300 & 23 \\
11 & 0 & 0 & 0 & 25 & -26 & 1 \\
12 & 0 & 0 & 0 & 1 & -1 & 0 \\ \hline
\end{tabular}

$N^{c=1}_{\twover,g,Q}$ for torus knot $(2,5)$
\end{center}  

\begin{center} 
\begin{tabular}{r|rrrr} \hline
g & Q=6 & 8 & 9 & 11 \\ \hline
0 & 66 & -330 & 330 & -66 \\
1 & 715 & -3905 & 3905 & -715 \\
2 & 3003 & -18656 & 18656 & -3003 \\
3 & 6435 & -47905 & 47905 & -6435 \\
4 & 8008 & -75218 & 75218 & -8008 \\
5 & 6188 & -77415 & 77415 & -6188 \\
6 & 3060 & -54248 & 54248 & -3060 \\
7 & 969 & -26333 & 26333 & -969 \\
8 & 190 & -8855 & 8855 & -190 \\
9 & 21 & -2024 & 2024 & -21 \\
10 & 1 & -300 & 300 & -1 \\
11 & 0 & -26 & 26 & 0 \\
12 & 0 & -1 & 1 & 0 \\ \hline
\end{tabular}

$N^{c=2}_{\twover,g,Q}$ for torus knot $(2,5)$
\end{center}  

\underline{Torus knot $(2,5)$ with framing $p=2$}

\begin{center}
\begin{tabular}{r|rrr} \hline
g & Q= 2 & 3 & 4 \\ \hline
0 & 3 & -3 & 1 \\
1 & 1 & -1 & 0 \\ \hline
\end{tabular}

$N^{c=1}_{\one,g,Q}$
\end{center} 

\begin{center} 
\begin{tabular}{r|rrrrrr} \hline
g & Q= 11/2 & 13/2 & 15/2 & 17/2 & 19/2 & 21/2 \\ \hline
0 & 175 & -585 & 580 & 28 & -330 & 132 \\
1 & 875 & -3480 & 3460 & 1554 & -3905 & 1496 \\
2 & 1820 & -9282 & 9261 & 10136 & -18656 & 6721 \\
3 & 2055 & -14384 & 14376 & 29985 & -47905 & 15873 \\
4 & 1377 & -14184 & 14183 & 51391 & -75218 & 22451 \\
5 & 561 & -9247 & 9247 & 56470 & -77415 & 20384 \\
6 & 136 & -4029 & 4029 & 41804 & -54248 & 12308 \\
7 & 18 & -1159 & 1159 & 21317 & -26333 & 4998 \\
8 & 1 & -211 & 211 & 7505 & -8855 & 1349 \\
9 & 0 & -22 & 22 & 1792 & -2024 & 232 \\
10 & 0 & -1 & 1 & 277 & -300 & 23 \\
11 & 0 & 0 & 0 & 25 & -26 & 1 \\
12 & 0 & 0 & 0 & 1 & -1	& 0 \\ \hline
\end{tabular}

$N^{c=1}_{\twohor,g,Q}$ for torus knot $(2,5)$
\end{center}  

\begin{center} 
\begin{tabular}{r|rrrr} \hline
g & Q=7 & 9 & 10 & 12 \\ \hline
0 & 66 & -330 & 330 & -66 \\
1 & 715 & -3905 & 3905 & -715 \\
2 & 3003 & -18656 & 18656 & -3003 \\
3 & 6435 & -47905 & 47905 & -6435 \\
4 & 8008 & -75218 & 75218 & -8008 \\
5 & 6188 & - 77415 & 77415 & -6188 \\
6 & 3060 & -54248 & 54248 & -3060 \\
7 & 969 & -26333 & 26333 & -969 \\
8 & 190 & -8855 & 8855 & -190 \\
9 & 21 & -2024 & 2024 & -21 \\
10 & 1 & -300 & 300 & -1 \\
11 & 0 & -26 & 26 & 0 \\
12 & 0 & -1 & 1 & 0 \\ \hline
\end{tabular}

$N^{c=2}_{\twohor,g,Q}$ for torus knot $(2,5)$
\end{center}  

\begin{center} 
\begin{tabular}{r|rrrrrr} \hline
g & Q= 11/2 & 13/2 & 15/2 & 17/2 & 19/2 & 21/2 \\ \hline
0 & 245 & -795 & 780 & 4 & -390 & 156 \\
1 & 1470 & -5545 & 5480 & 1910 & -5395 & 2080 \\
2 & 3724 & -17444 & 17361 & 15456 & -30108 & 11011 \\
3 & 5215 & -32075 & 32030 & 54461 & -90376 & 30745 \\
4 & 4445 & -37932 & 37921 & 110395 & -166595 & 51766 \\
5 & 2394 & -30178 & 30177 & 143961 & -202930 & 56576 \\
6 & 817 & -16472 & 16472 & 127771 & -170408 & 41820 \\
7 & 171 & -6175 & 6175 & 79440 & -100929 & 21318 \\
8 & 20 & -1561 & 1561 & 34978 & -42503 & 7505 \\
9 & 1 & -254 & 254 & 10857 & -12650 & 1792 \\
10 & 0 & -24 & 24 & 2323 & -2600 & 277 \\
11 & 0 & -1 & 1 & 326 & -28 & 25 \\
12 & 0 & 0 & 0 & 27 & -1 & 1 \\
13 & 0 & 0 & 0 & 1 & 0 & 0 \\ \hline
\end{tabular}

$N^{c=1}_{\twover,g,Q}$ for torus knot $(2,5)$
\end{center}  

\begin{center} 
\begin{tabular}{r|rrrr} \hline
g & Q=7 & 9 & 10 & 12 \\ \hline
0 & 78 & -390 & 390 & -78 \\
1 & 1001 & -5395 & 5395 & -1001 \\
2 & 5005 & -30108 & 30108 & -5005 \\
3 & 12870 & -90376 & 90376 & -12870 \\
4 & 19448 & -166595 & 166595 & -19448 \\
5 & 18564 & -202930 & 202930 & -18564 \\
6 & 11628 & -170408 & 170408 & -11628 \\
7 & 4845 & -100929 & 100929 & -4845 \\
8 & 1330 & -42503 & 42503 & -1330 \\
9 & 231 & -12650 & 12650 & -231 \\
10 & 23 & -2600 & 2600 & -23 \\
11 & 1 & -351 & 351 & -1 \\
12 & 0 & -28 & 28 & 0 \\
13 & 0 & -1 & 1 & 0 \\ \hline
\end{tabular}

$N^{c=2}_{\twover,g,Q}$ for torus knot $(2,5)$
\end{center}

\underline{Connected sum of trefoil and trefoil with framing $p=0$}

\begin{center}
\begin{tabular}{r|rrrr}\hline
g & Q=2 & 3 & 4 & 5 \\ \hline
0 & 8 & -14 & 9 & -2 \\
1 & 6 & -11 & 6 & -1 \\
2 & 1 & -2 & 1 & 0 \\ \hline
\end{tabular}

$N_{\one,g,Q}^{c=1}$ for trefoil \# trefoil
\end{center}

\begin{center}
\begin{tabular}{r|rrrrrrrr} \hline
g & Q=7/2 & 9/2 & 11/2 & 13/2 & 15/2 & 17/2 & 19/2 & 21/2 \\ \hline
0 & 143 & -831 & 1950 & -2366 & 1561 & -525 & 66 & 2\\
1 & 404 & -3144 & 8854 & -11819 & 7544 & -1488 & -596 & 245 \\
2 & 464 & -5419 & 19211 & -28097 & 14046 & 6348 & -9194 & 2641 \\
3 & 277 & -5379 & 25184 & -40255 & 6296 & 44160 & -41756 & 11473 \\
4 & 90 & -3292 & 21666 & -38551 & -18588 & 110890 & -98450 & 26235 \\
5 & 15 & -1256 & 12654 & -26241 & -38613 & 159091 & -141400 & 35750 \\
6 & 1 & -290 & 5048 & -13093 & -36589 & 147270 & -133378 & 31031 \\
7 & 0 & -37 & 1352 & -4787 & -21053 & 92681 & -85919 & 17763 \\
8 & 0 & -2 & 232 & -1243 & -7860 & 40544 & -38455 & 6784 \\
9 & 0 & 0 & 23 & -215 & -1917 & 12353 & -11954 & 1710 \\
10 & 0 & 0 & 1 & -22 & -295 & 2574 & -2531 & 273 \\
11 & 0 & 0 & 0 & -1 & -26 & 350 & -348 & 25 \\
12 & 0 & 0 & 0 & 0 & -1 & 28 & -28 & 1 \\
13 & 0 & 0 & 0 & 0 & 0 & 1 & -1 & 0 \\ \hline
\end{tabular}

$N_{\twohor,g,Q}^{c=1}$ for trefoil \# trefoil
\end{center}

\begin{center}
\begin{tabular}{r|rrrrrrrr} \hline
g & Q=4 & 5 & 6 & 7 & 8 & 9 & 10 & 11 \\ \hline
0 & -46 & 627 & -2210 & 3524 & -2891 & 1190 & -193 & -1 \\
1 & -115 & 2857 & -12709 & 23835 & -22244 & 10068 & -1517 & -121 \\
2 & -114 & 5764 & -32974 & 73721 & -78050 & 37511 & -4648 & -1210 \\
3 & -54 & 6412 & -48952 & 133320 & -160443 & 79607 & -5171 & -4719 \\
4 & -12 & 4241 & -45575 & 155369 & -214257 & 107758 & 1914 & -9438 \\
5 & -1 & 1707 & -27770 & 122272 & -195972 & 98868 & 11907 & -11011 \\
6 & 0 & 410 & -11234 & 66279 & -126105 & 63513 & 15145 & -8008 \\
7 & 0 & 54 & -2987 & 24753 & -57626 & 28938 & 10608 & -3740 \\
8 & 0 & 3 & -501 & 6247 & -18593 & 9314 & 4652 & -1122 \\
9 & 0 & 0 & -48 & 1016 & -4138 & 2070 & 1309 & -209 \\
10 & 0 & 0 & -2 & 96 & -604 & 302 & 230 & -22 \\
11 & 0 & 0 & 0 & 4 & -52 & 26 & 23 & -1 \\
12 & 0 & 0 & 0 & 0 & -2 & 1 & 1 & 0 \\ \hline
\end{tabular}

$N^{c=2}_{\twohor,g,Q}$ for trefoil \# trefoil
\end{center}

\begin{center}
\begin{tabular}{r|rrrrrrrr} \hline
g & Q=7/2 & 9/2 & 11/2 & 13/2 & 15/2 & 17/2 & 19/2 & 21/2 \\ \hline
0 & 227 & -1237 & 2756 & -3206 & 2045 & -671 & 84 & 2 \\
1 & 801 & -5621 & 14872 & -19187 & 12269 & -2861 & -564 & 291 \\
2 & 1190 & -11771 & 38341 & -54346 & 29773 & 5595 & -12485 & 3697 \\
3 & 955 & -14403 & 59796 & -92245 & 27489 & 67819 & -68353 & 18942 \\
4 & 444 & -11132 & 61614 & -104212 & -18925 & 211571 & -190697 & 51337 \\
5 & 119 & -5578 & 43750 & -83517 & -79482 & 364896 & -323843 & 83655 \\
6 & 17 & -1803 & 21761 & -49317 & -100043 & 404876 & -363462 & 87971 \\
7 & 1 & -362 & 7561 & -21758 & -73516 & 307780 & -281842 & 62136 \\
8 & 0 & -41 & 1795 & -7097 & -35330 & 165164 & -154547 & 30056 \\
9 & 0 & -2 & 277 & -1653 & -11446 & 63250 & -60402 & 9976 \\
10 & 0 & 0 & 25 & -258 & -2483 & 17202 & -16719 & 2233 \\
11 & 0 & 0 & 1 & -24 & -346 & 3248 & -3201 & 322 \\
12 & 0 & 0 & 0 & -1 & -28 & 405 & -403 & 27 \\
13 & 0 & 0 & 0 & 0 & -1 & 30 & -30 & 1 \\
14 & 0 & 0 & 0 & 0 & 0 & 1 & -1 & 0 \\ \hline
\end{tabular}

$N_{\twover,g,Q^{c=1}}$ for trefoil \# trefoil
\end{center}

\begin{center}
\begin{tabular}{r|rrrrrrrr} \hline
g & Q= 4 & 5 & 6 & 7 & 8 & 9 & 10 & 11 \\ \hline
0 & -74 & 869 & -2921 & 4526 & -3631 & 1466 & -234 & -1 \\
1 & -234 & 4781 & -20075 & 36309 & -32999 & 14673 & -2311 & -144 \\
2 & -310 & 11808 & -62281 & 132252 & -135333 & 64116 & -8536 & -1716 \\
3 & -212 & 16343 & -110933 & 280964 & -323696 & 159083 & -13541 & -8008 \\
4 & -77 & 13779 & -125127 & 386321 & -503851 & 252187 & -3927 & -19305 \\
5 & -14 & 7338 & -93938 & 362625 & -540953 & 272586 & 19812 & -27456 \\
6 & -1 & 2479 & -48041 & 238667 & -413588 & 208502 & 36734 & -24752 \\
7 & 0 & 515 & -16793 & 111146 & -228664 & 115027 & 33457 & -14688 \\
8 & 0 & 60 & -3945 & 36403 & -91650 & 45984 & 18962 & -5814 \\
9 & 0 & 3 & -595 & 8191 & -26359 & 13199 & 7081 & -1520 \\
10 & 0 & 0 & -52 & 1204 & -5298 & 2650 & 1748 & -252 \\
11 & 0 & 0 & -2 & 104 & -706 & 353 & 275 & -24 \\
12 & 0 & 0 & 0 & 4 & -56 & 28 & 25 & -1 \\
13 & 0 & 0 & 0 & 0 & -2 & 1 & 1 & 0 \\ \hline
\end{tabular}

$N^{c=2}_{\twover,g,Q}$ for trefoil \# trefoil
\end{center}

\subsection{Framed Links}

We take Hopf Link $H(p_1,p_2)$ with linking number -1 and framing on the two
component knots as $p_1$ and $p_2$. The integers $N_{(R_1,R_2),g,Q}^{c=1}$
and $N_{(R_1,R_2),g,Q}^{c=2}$
for various combinations of $p_1$ and $p_2$ are tabulated below.\\

\underline{$p_1=0=p_2$}

$$N^{c=1}_{\one,\one,0,\pm 1/2}=\pm 1$$

\begin{center}
\begin{tabular}{r|rr} \hline
g & Q=-1 & 0 \\ \hline
0 & -1 & 1 \\ \hline
\end{tabular}
\hspace{.6in}
\begin{tabular}{r|rr} \hline
g & Q=0 & 1 \\ \hline
0 & 1 & -1 \\ \hline
\end{tabular}

$N^{c=1}_{\twohor,\one,g,Q}$ for hopf link \hspace{.2in}
$N^{c=1}_{\twover,\one,g,Q}$ for hopf link
\end{center}

\underline{$p1=1=p2$}

\begin{center} 
\begin{tabular}{r|rr} \hline
g & Q=1/2 & 3/2 \\ \hline
0 & -1 & 1 \\ \hline
\end{tabular}
\hspace{.6in}
\begin{tabular}{r|rr} \hline
g & Q=3/2 & 5/2 \\ \hline
0 & -1 & 1 \\ \hline
\end{tabular}
\hspace{.6in}
\begin{tabular}{r|rrr} \hline
g & Q=1/2 & 3/2 & 5/2 \\ \hline
0 & 1 & -5 & 4 \\
1 & 0 & -1 & 1 \\ \hline
\end{tabular}

$N^{c=1}_{\one,\one,g,Q}$ for hopf link \hspace{.2in}
$N^{c=1}_{\twohor,\one,g,Q}$ for hopf link \hspace{.2in}
$N^{c=1}_{\twover,\one,g,Q}$ for hopf link
\end{center} 

\underline{$p1=2=p2$}

\begin{center} 
\begin{tabular}{r|rr} \hline
g & Q=3/2 & 5/2 \\ \hline
0 & -1 & 1 \\ \hline
\end{tabular}
\hspace{.6in}
\begin{tabular}{r|rrr} \hline
g & Q=2 & 3 & 4 \\ \hline
0 & -1 & 5 & -4 \\
1 & 0 & 1 & -1 \\ \hline
\end{tabular}
\hspace{.6in}
\begin{tabular}{r|rrr} \hline
g & Q=2 & 3 & 4 \\ \hline
0 & -4 & 13 & -9 \\
1 & -1 & 7 & -6 \\ 
2 & 0 & 1 & -1 \\ \hline
\end{tabular}

$N^{c=1}_{\one,\one,g,Q}$ for hopf link \hspace{.2in}
$N^{c=1}_{\twohor,\one,g,Q}$ for hopf link \hspace{.2in}
$N^{c=1}_{\twover,\one,g,Q}$ for hopf link
\end{center} 

\underline{$p1=2$, $p2=3$}

\begin{center} 
\begin{tabular}{r|rr} \hline
g & Q=2 & 3 \\ \hline
0 & 1 & -1 \\ \hline
\end{tabular}
\hspace{.5in}
\begin{tabular}{r|rrr} \hline
g & Q=5/2 & 7/2 & 9/2 \\ \hline
0 & 1 & -5 & 4 \\
1 & 0 & -1 & 1 \\ \hline
\end{tabular}
\hspace{.6in}
\begin{tabular}{r|rrr} \hline
g & Q=5/2 & 7/2 & 9/2 \\ \hline
0 & 4 & -13 & 9 \\
1 & 1 & -7 & 6 \\ 
2 & 0 & -1 & 1 \\ \hline
\end{tabular}
$N^{c=1}_{\one,\one,g,Q}$ for hopf link \hspace{.5in}
$N^{c=1}_{\twohor,\one,g,Q}$ for hopf link \hspace{.4in}
$N^{c=1}_{\twover,\one,g,Q}$ for hopf link
\end{center} 
\section{Summary and Discussions}
We have explicitly demonstrated the evaluation of
framed knot and link invariants carrying composite
representations in $U(N)$ Chern-Simons gauge theory.
Particularly, we argued a specific choice for the $U(1)$ charge
corresponding to the composite representations (\ref {framchar}) so that 
the composite invariants for framed knots and links are
polynomials in variables $q,\lambda$.
Further, this direct method enabled us to verify
generalised Rudolph's theorem for many framed knots (\ref {rud}).

\noindent
The composite invariants was very essential to obtain the untwisted
sector open topological string amplitude(\ref {con1a}) 
on the orientifold of the resolved conifold geometry. 
Similar to Ooguri-Vafa conjecture (\ref {unrfm}),
Marino\cite {mar9} conjectured a form for the
reformulated invariants (\ref {horg}) and the 
refined reformulated invariants (\ref {hath}).
We have verified the conjecture for many framed knots and links
and presented the reformulated invariants for few examples.
The cross-cap $c=0$ BPS integer coefficients (\ref {hath}) 
are also tabulated for these examples.

In earlier works\cite {vinc1,vinc2,prav}, there was  difficulty 
in seperating $c=0$ and $c=2$
contribution from  the topological string free energy(\ref {orie1}) 
but using the parity argument in variable $\sqrt{\lambda}$, 
the cross-cap $c=1$ amplitude could be
determined. With the present work on composite
invariants following the approach \cite {mar9},
we can determine the unoriented topological string amplitude (\ref {orie1}) 
by subtracting the untwisted sector
contribution from the free energy of the open topological string theory on
the orientifold. We have checked that the reformulated
$SO$ invariants obtained from the unoriented topological string free
energy also obeys Marino's conjectured
form (\ref {horg}). Further, the refined $SO$ reformulated invariants
obtained using eqn. (\ref {reg}) satisfies the conjectured form(\ref {hatg}).
We have tabulated the BPS integer invariants corresponding to 
cross-cap $c=1$ and $c=2$ obtained from 
reformulated invariants (\ref {hatg}) for some framed knots
and links. In particular, the $c=1$ integer coefficients agrees
with our earlier work \cite {prav}. Also, the
BPS integer coefficients for the standard framing ($p=0$) torus knots and 
torus links agrees with the results in Ref.\cite {stev}.
The verification of Marino's conjectures for many framed knots and 
two-component framed links indirectly confirms that our choice of the 
$U(1)$ charge (\ref {framchar}) for the composite representations is correct.

The Marino's conjectures, which we verified for some torus knots and 
torus links, should be obeyed by non-torus knots and non-torus links 
as well. The Chern-Simons approach requires the 
$SU(N)$ quantum Racah coefficients for the non-torus knot 
invariant evaluation. Unfortunately,
these coefficients are not available in the literature.
In Ref.\cite {rama}, the $SU(N)$ quantum Racah coefficients for
some representations could be determined using isotopy 
equivalence of knots enabling evaluation of non-torus knot 
invariants. We believe that there must be a similar
approach of determining composite invariants for the non-torus
knots.

It will be interesting to generalise these integrality properties
in the context of Khovanov homology \cite {gukov} and
Kauffman homology \cite {gukov1}. We hope
to report on this work in a future publication.

\vskip.2cm

\begin{center}
{\bf Acknowledgments}
\end{center}
CP would like to express her deep sense of gratitude to 
Prof. Raghava Varma for giving encouragement and providing all the 
necessary help during the course of this work.She is also very much thankful 
to Dr.Ankhi Roy for her sincere cooperation which helped her to 
complete this work. PR would like to thank IRCC, IIT Bombay
for the research grant. CP would like to thank
the organisers for giving her the opportunity
to present this work in  the national strings meeting 
(NSM) held at IIT Bombay during Feb 10-15, 2010. 

\newpage
\vskip.5cm

\appendix {\bf {\Large{Appendix}}} 
\section{$U(N)$ Composite Knot Invariants}
\subsection{For (R,S)=(2 horizonal box, single box)}
Let us denote the composite representation $(\twohor,\one)$ by $\rho_{02}$
whose highest weight is 
$\Lambda^{(N-1)}+2\Lambda^{(1)}.$
The highest weights, quantum dimensions and
the braiding eigenvalues corresponding to the irreducible representations
$R_t \in \rho_{02}\otimes\rho_{02}$ are  
{\small
\begin{displaymath}
\begin{array}{|l|l|l|l|}\hline
R_t&{\rm highest~ weight}&{\rm quantum~ dimension}&\lambda_t\\ \hline
~&~&~&~\\
R_1 & 2\Lambda^{(N-1)}+4\Lambda^{(1)} &
\frac{[N]^2[N-1][N+1][N+2][N+5]}{[4][3][2][2]}& q^{-3/2} \lambda^{3/2} \\
~&~&~&~\\
R_2 & \Lambda^{(N)}+{\Lambda^{(N-2)}}+4\Lambda^{(1)}&
\frac{[N][N-1][N-2][N+1][N+3][N+4}{[4][3][2][2]}& - q^{-1/2} \lambda^{3/2}  \\
~&~&~&\\
R_3 & 
2\Lambda^{(N-1)}+{\Lambda^{(2)}}+2\Lambda^{(1)}&
\frac{[N][N-1][N-2][N+1]^2[N+4]}{[4][2][2]} & -q^{1/2} \lambda^{3/2} \\
~&~&~&\\
R_4 & \Lambda^{(N)}+{\Lambda^{(N-2)}}+\Lambda^{(2)}+2\Lambda^{(1)}&
\frac{[N]^2[N-1][N+1][N+3]}{[4][2][2]}& q^{3/2} \lambda^{3/2} \\
~&~&~&\\
R_5 & \Lambda^{(N)}+\Lambda^{(N-1)}+3\Lambda^{(1)} & 
\frac{[N][N-1][N+1][N+3]}{[3][2]} & q^{1/2} \lambda^2 \\
~&~&~&\\
R_6 &\Lambda^{(N)}+\Lambda^{(N-1)}+\Lambda^{(2)}+{\Lambda^{(1)}}&
 \frac{[N]^2[N-2][N+2]}{[3]}&  q^2 \lambda^2  \\
~&~&~&\\
R_7 & 2\Lambda^{(N-1)}+2\Lambda^{(2)}& 
\frac{[N][N-1]^2[N-2][N+2][N+3]}{[3][2][2][2]}&  q^{3/2} \lambda^{3/2}  \\
~&~&~&\\
R_8 &
\Lambda^{(N)}+{\Lambda^{(N-2)}}+2\Lambda^{(2)}&
\frac{[N][N-2][N-3][N+1]^2[N+2]}{[3][2][2][2]}&  -q^{5/2} \lambda^{3/2} \\
%&~&~&\\
R_9& 2\Lambda^{(N)}+2\Lambda^{(1)} &
\frac{[N][N+1]}{[2]}& q^{3/2} \lambda^{5/2} \lambda \\
~&~&~&\\
R_{10} & 2\Lambda^{(N)}+\Lambda^{(2)}&
\frac{[N][N-1]}{[2]}& -q^{5/2} \lambda^{5/2}  \\
~&~&~&\\
R_{11} & \Lambda^{(N)}+\Lambda^{(N-1)}+3\Lambda^{(1)} &
 \frac{[N][N-1][N+1][N+3]}{[3][2]}& -q^{1/2}\lambda^2 \\
~&~&~&\\
R_{12} & 
\Lambda^{(N)}+\Lambda^{(N-1)}+\Lambda^{(2)}+{\Lambda^{(1)}}&
\frac{[N]^2[N-2][N+2]}{[3]}&  -q^2 \lambda^2 \\ 
~&~&~&\\ \hline
 \end{array}
\end{displaymath}
}
Using the above table, we can evaluate directly the composite invariants
for knots obtained as closure of two strand braids. We present
few composite invariants for framed trefoil, framed five-crossing knot $5_1$ 
and framed seven-crossing knot $7_1$ with arbitrary framing $p$:
{\small
\begin{eqnarray}
{\cal H}_{(\twohor,\one)}[{\cal K}_3]&=&(dim_q \rho_{02})q^{-3}{\lambda^{\tp2}}\left({\lambda^3}\right.\nonumber\\
&&(q^{8}+2 q^{6}+q^{5}+q^4+q^3+q^2+1) \nonumber \\
&&+ {\lambda^4}(-1-q-2 q^{3}-2 q^{4}-q^{5}-2q^{6}-q^{7}-2 q^{9})\nonumber \\
&&+ {\lambda^{5}}(q+2 q^{4}+q^{5}-q^{6}+2 q^{7}+q^{10})
+ {\lambda^{6}}(-q^{5}+q^{6}-2 q^{8}+q^{9})\nonumber \\
&& \left. + {\lambda^{7}}(-q^{7}+q^{8}+q^{9}-q^{10})\right)\\ 
{\cal H}_{(\twohor,\one)}[{\cal K}_5]&=&(dim_q \rho_{02})q^{-6}{\lambda^{\tp2}}\left( {\lambda^{6}}
(1+q^2+q^3+2q^4+q^5+2q^6+2q^7+3q^8+2q^9\right.\nonumber\\
&&+3q^{10}+2q^{11}+3q^{12} +q^{13}+2q^{14}+q^{16})
+ {\lambda^{7}}(- 1-q-q^{2}-2 q^{3}-2 q^{4}-4 q^{5}\nonumber\\
&&-3 q^{6} -4 q^{7} - 6q^{8} -5 q^{9}-4 q^{10}-6 q^{11}-4 q^{12}-3q^{13}-3 q^{14}
-3 q^{15}-2 q^{17})\nonumber \\
&&+ {\lambda^{8}}(q+q^{3}+q^{4}+2q^{5}+3 q^{6}+2 q^{7}+3 q^{8}
+ 6 q^{9}+4 q^{11}+2q^{13}+3 q^{15}+q^{10}\nonumber \\
&& +4 q^{12}+2 q^{14}
+q^{18})
+ {\lambda^{9}}(-q^{6}-q^{7}+q^{8}-2 q^{9}-2 q^{10}+2 q^{11}
-q^{13}
+ q^{14}\nonumber \\
&& -q^{16}+q^{17}-4 q^{12}-q^{15})
+ {\lambda^{10}}(-q^{9}+2q^{10}-q^{11}+3 q^{13}-2 q^{14}-q^{15}
-q^{12}+q^{16})\nonumber \\
&& \left. + {\lambda^{11}}(-q^{11}+2q^{12}-q^{13}-q^{14}-q^{16}+2 q^{15})
+ {\lambda^{12}}(-q^{13}+q^{14}+q^{17}-q^{18}) \right)\\ 
{\cal H}_{(\twohor,\one)}[{\cal K}_7]&=&(dim_q \rho_{02}){q^{-9}}{\lambda^{\tp2}}\left( {\lambda^{9}}(1+q^2+q^3+2 q^4+q^5+3q^6\right. \nonumber\\
&&+2q^7+3q^8+3q^9+4q^{10}+3q^{11} +5q^{12} \nonumber \\
&&+4q^{13}+5q^{14}+4q^{15}+5q^{16}+3q^{17}+5q^{18}+2q^{19}+3q^{20}+q^{21}+2q^{22}+q^{24})\nonumber \\
&& + {\lambda^{10}}(-1-q-q^{2}-2 q^{3}-3 q^{4}-4 q^{5}-3 q^{6}-6 q^{7}
- 6 q^{8}-7 q^{9}-8 q^{10}-9q^{11}\nonumber \\
&& -9 q^{12}-11 q^{13}-10 q^{14}-10 q^{15}-10 q^{16}
- 10q^{17}-7 q^{18}-8 q^{19}-5 q^{20}-5 q^{21}\nonumber \\
&& -3 q^{22}-3q^{23}-2 q^{25})
+ {\lambda^{11}}(q+q^{3}+q^{4}+3 q^{5}+2 q^{6}+3 q^{7}
+5 q^{8}+ 5 q^{9}+5 q^{10}\nonumber \\
&& +8 q^{11}+6 q^{12}+9 q^{13}+8 q^{14}+8 q^{15}+7 q^{16}+10 q^{17}+5 q^{18}+6 q^{19}
+6 q^{20}+3 q^{21}\nonumber \\
&& + 2 q^{22}+3q^{23}+q^{26})
+ {\lambda^{12}}(-q^{6}-q^{8}-2 q^{9}-2 q^{11}-3 q^{12}-4 q^{14}-2 q^{15}-2 q^{16}\nonumber \\
&& -
3 q^{17}-4 q^{18}+ q^{19}-3 q^{20}- q^{21}+q^{22}-q^{23}-q^{24}+q^{25})\nonumber \\
&& + {\lambda^{13}}(-q^{11}+2 q^{12}-q^{13}-q^{14}+2 q^{15}-q^{16}+q^{18}-q^{19}-q^{20}+2 q^{21}
-q^{22}\nonumber \\
&& - q^{23}+q^{24})
+{\lambda^{14}} (-q^{13}+2 q^{14}-q^{15}-q^{16}+2 q^{17}-q^{18}+q^{20}-q^{21}-q^{22}+ q^{23})\nonumber \\
&& + {\lambda^{15}}(-q^{15}+2 q^{16}-q^{17}-q^{18}+2 q^{19}-q^{20})
+{\lambda^{16}} (-q^{17}+2 q^{18}-q^{19}-q^{20}\nonumber \\
&& \left. +q^{21}+q^{23}-q^{24} )
+ {\lambda^{17}}(-q^{19}+q^{20}+q^{25}-q^{26}) \right)
\end{eqnarray}
}
\subsection{(R,S)=(two vertical box, single box)}
Let us denote the composite representation $(\twover,\one)$ by $\rho_{03}$
and its highest weight is $\rho_{03} = {\Lambda^{(N-1)}}+{\Lambda^{(2)}}~.$
The representations $R_t$ obtained from $\rho_{03}\otimes\rho_{03}$
and their quantum dimensions and the signs of the braiding eigenvalues: 
$\epsilon_t$ are 
{\small
\begin{displaymath}
\begin{array}{ll}
R_1 = 2\Lambda^{(N-1)}+2\Lambda^{(2)}; \epsilon_1= 1 &
R_2 = \Lambda^{(N)}+{\Lambda^{(N-2)}}+2\Lambda^{(2)};\epsilon_2= -1\\
R_3 = \Lambda^{(N)}+\Lambda^{(N-1)}+{\Lambda^{(2)}}+\Lambda^{(1)};
\epsilon_3 =1 \qquad&
R_4 = 2\Lambda^{(N)}+2\Lambda^{(1)};\epsilon_4 =1\\
R_5 = 2\Lambda^{(N-1)}+\Lambda^{(3)}+\Lambda^{(1)};\epsilon_5 =-1 &
R_6 = \Lambda^{(N)}+\Lambda^{(N-2)}+\Lambda^{(3)}+{\Lambda^{(1)}};\epsilon_6 =1\\
R_7 = 2\Lambda^{(N)}+\Lambda^{(2)}; \epsilon_7 = -1&
R_8 = \Lambda^{(N)}+{\Lambda^{(N-1)}}+\Lambda^{(3)};\epsilon_8 =1\\
R_9 = \Lambda^{(N-1)}+\Lambda^{(N-3)}+\Lambda^{(4)};\epsilon_9 =1 &
R_{10} = \Lambda^{(N-1)}+\Lambda^{(N-3)}+\Lambda^{(4)}; \epsilon_{10} =-1\\
R_{11} = \Lambda^{(N)}+\Lambda^{(N-1)}+\Lambda^{(2)}+\Lambda^{(1)}; \epsilon_{11} = -1&
R_{12} = \Lambda^{(N)}+\Lambda^{(N-1)}+\Lambda^{(3)};\epsilon_{12} = -1
\end{array}
\end{displaymath}
}
For the above irreducible representations, quadratic casimir and
quantum dimensions can be computed using eqns.(\ref{casi},\ref{egval}).
With this data, the polynomials of the framed knots carrying the 
composite representation can be computed. The  composite
invariants of some of the $p$-framed torus knots of type $(2,2m+1)$ 
are:
{\small
\begin{eqnarray}
{\cal H}_{(\twover,\one)}[{\cal K}_3]&=&(dim_q \rho_{03}){q^{-7}}{\lambda^{\tp2}}\left({\lambda^3}(q^2+2 q^4+q^{5}+q^{6}+q^{7}+q^{8}+q^{10})\right.\nonumber \\
&& +{\lambda^4}(-2 q-q^{3}-2 q^{4}-q^{5}-2q^{6}-2 q^{7}-q^{9}-q^{10})\nonumber \\
&& + {\lambda^{5}}(1+2 q^{3}-q^{4}+q^{5}+2 q^{6}+q^{9})\nonumber \\
&&\left. + {\lambda^{6}}(q-2 q^{2}+q^{4}-q^{5})
+ {\lambda^{7}}(-1+q+q^{2}-q^{3})\right)
\end{eqnarray}
\begin{eqnarray}
{\cal H}_{(\twover,\one)}[{\cal K}_5]&=&(dim_q \rho_{03}){q^{-12}}{\lambda^{\tp2}}\left( {\lambda^{6}}(q^2+2 q^4+q^{5}+3 q^{6}+2 q^{7}+3 q^{8}+2q^{9}+ 3 q^{10}
+2 q^{11}+2 q^{12}\right. \nonumber\\
&&+q^{13} +2 q^{14}+q^{15}+q^{16}+q^{18})
+ {\lambda^{7}}(-2 q-3 q^{3} -3 q^{4}-3 q^{5}
-4 q^{6}-6 q^{7}-4 q^{8}-5 q^{9}\nonumber \\
&& -6 q^{10}-4 q^{11}-3 q^{12}
-4 q^{13}-2q^{14}
- 2 q^{15}-q^{16}-q^{17}-q^{18} )\nonumber \\
&& + {\lambda^{8}}(1+3 q^{3}+2 q^{4}+2q^{5}+4 q^{6}+4 q^{7}+q^{8}
+6 q^{9}+3 q^{10}
+ 2 q^{11}+3 q^{12}+2q^{13}\nonumber \\
&& +q^{14}+q^{15}+q^{17}) 
+ {\lambda^{9}}(q-q^{2}-q^{3}+q^{4}-q^{5}-4 q^{6}+2 q^{7}
- 2 q^{8}-2 q^{9}+q^{10}\nonumber \\
&& -q^{11}-q^{12})
+ {\lambda^{10}}(q^{2}-q^{3}-2 q^{4}-q^{6}-q^{7}+2 q^{8}
-q^{9}+3 q^{10})\nonumber \\
&&\left. + {\lambda^{11}}(-q^{2}+2 q^{3}-q^{4}-q^{5}+2 q^{6}-q^{7})
+ {\lambda^{12}}(-1+q+q^{4}-q^{5})\right)
\end{eqnarray}
\begin{eqnarray}
{\cal H}_{(\twover,\one)}[{\cal K}_7]&=&(dim_q \rho_{03}){q^{-17}}{\lambda^{\tp2}}\left( {\lambda^{9}} ( q^2+2 q^4+q^{5}+3 q^{6}+2 q^{7}+5 q^{8}
+ 3q^{9}+5 q^{10}+4 q^{11}+5 q^{12}\right.\nonumber\\
&&+4 q^{13} +5 q^{14}+3 q^{15}+4 q^{16}+ 3q^{17}
+3 q^{18}+2 q^{19}+3 q^{20}+q^{21}+2 q^{22}+q^{23}+q^{24}+q^{26})\nonumber \\
&& +{\lambda^{10}}(-3 q^{3}-3 q^{4}-5 q^{5}-5q^{6}-2 q-8 q^{7}-7 q^{8}-10 q^{9}
-10 q^{10}-10 q^{11}-10 q^{12}\nonumber \\
&& -11q^{13}-9 q^{14}-9 q^{15}-8 q^{16}-7 q^{17}
- 6 q^{18}-6q^{19}-3 q^{20}-4 q^{21}-3 q^{22}-2q^{23}\nonumber \\
&& -q^{24}-q^{25}
- q^{26})
+{\lambda^{11}} (1+ 3 q^{3}+2 q^{4}+3 q^{5}+6q^{6}+6 q^{7}+5 q^{8}+10 q^{9}+7 q^{10}\nonumber \\
&& +8 q^{11}
+8 q^{12}+ 9 q^{13}+6 q^{14}+8q^{15}+5 q^{16}+5 q^{17}+5 q^{18}+3 q^{19}+2 q^{20}+3 q^{21}\nonumber \\
&& +q^{22}+q^{23}
+ q^{25})
+{\lambda^{12}} ( q-q^{2}-q^{3}+q^{4}-q^{5}-3 q^{6}+q^{7}
-4 q^{8}
- 3 q^{9}-2 q^{10}\nonumber \\
&& -2 q^{11}-4 q^{12}-3 q^{14}-2 q^{15}-2 q^{17}-q^{18}-q^{20})
+{\lambda^{13}} ( q^{2}-q^{3}-q^{4}+2q^{5}-q^{6}\nonumber \\
&& -q^{7}+q^{8}
-q^{10}
+ 2 q^{11}-q^{12}-q^{13}+2 q^{14}-q^{15})
+{\lambda^{14}} ( q^{3}-q^{4}-q^{5}+q^{6}-q^{8}\nonumber \\
&& +2 q^{9}-q^{10}-q^{11}
+2 q^{12}
- q^{13})
-{\lambda^{15}} ( q^{6}-2 q^{7}+q^{8}+q^{9}-2 q^{10}+q^{11})\nonumber \\
&&\left. -{\lambda^{16}} ( +q^{2}-q^{3}-q^{5}+q^{6}+q^{7}-2 q^{8}+q^{9})
+{\lambda^{17}} ( -1+q+q^{6}-q^{7})\right)
\end{eqnarray}
}
\subsection{ (R,S)=(two horizontal box, two horizontal box)}
Let us denote the composite representation $(\twohor,\twohor)$ by $\rho_{04}$
whose highest weight is  
\begin{equation}
\rho_{04} = 2{\Lambda^{(N-1)}}+2{\Lambda^{(1)}}
\end{equation}
The highest weights of the ireducible representations $R_t$
 obtained from $\rho_{04}\otimes\rho_{04}$ 
and the signs of the braiding eigenvalues $\epsilon_t$ are
\vskip.1cm
{\small
\begin{tabular}{|l|l|l||l|l|l|}\hline
$R_t$&Highest weight&$\epsilon_t$&$R_t$&Highest weight&$\epsilon_t$\\
\hline
$R_1$&$4\Lambda^{(N-1)}+4\Lambda^{(1)}$&1& $R_2$&$\Lambda^{(N)}
+2\Lambda^{(N-1)}+{\Lambda^{(N-2)}}+4\Lambda^{(1)}$&-1 \\
$R_3$&$2{\Lambda^{(N)}}+2{\Lambda^{(N-2)}}+4{\Lambda^{(1)}}$&1 &
$R_4$&${\Lambda^{(N)}}+3{\Lambda^{(N-1)}}+3{\Lambda^{(1)}}$& 1\\
$R_5$& $2{\Lambda^{(N)}}+{\Lambda^{(N-1)}}+{\Lambda^{(N-2)}}
+3{\Lambda^{(1)}}$& 1&$R_6$&${\Lambda^{(N)}}+2{\Lambda^{(N-1)}}+
{\Lambda^{(N-2)}}+ {\Lambda^{(2)}} + 2{\Lambda^{(1)}}$&1\\
$R_7$&$ 2{\Lambda^{(N)}}+2{\Lambda^{(N-2)}}+{\Lambda^{(2)}}+
2{\Lambda^{(1)}}$&-1&
$R_8$&$2{\Lambda^{(N)}}+2{\Lambda^{(N-1)}}+2{\Lambda^{(1)}}$&1\\
$R_9$& $2{\Lambda^{(N)}}+{\Lambda^{(N-1)}}+{\Lambda^{(N-2)}}+
{\Lambda^{(2)}}+{\Lambda^{(1)}}$&-1&
$R_{10}$&$2{\Lambda^{(N)}}+2{\Lambda^{(N-2)}}+2{\Lambda^{(2)}}$&1\\
$R_{11}$& $2{\Lambda^{(N)}}+2{\Lambda^{(N-1)}}+{\Lambda^{(2)}}$& -1&
$R_{12}$&$3{\Lambda^{(N)}}+{\Lambda^{(N-2)}}+{\Lambda^{(2)}}$& 1\\
$R_{13}$&$3{\Lambda^{(N)}}+{\Lambda^{(N-1)}}+{\Lambda^{(1)}}$& 1&
$R_{14}$&$4{\Lambda^{(N)}}$& 1\\
$R_{15}$&$4{\Lambda^{(N-1)}}+{\Lambda^{(2)}}+2{\Lambda^{(1)}}$& -1&
$R_{16}$& $4{\Lambda^{(N-1)}}+2{\Lambda^{(2)}}$& 1\\
$R_{17}$& ${\Lambda^{(N)}}+3{\Lambda^{(N-1)}}+3{\Lambda^{(1)}}$&-1&
$R_{18}$&$2{\Lambda^{(N)}}+{\Lambda^{(N-1)}}+{\Lambda^{(N-2)}}+
3{\Lambda^{(1)}}$& -1\\
$R_{19}$&${\Lambda^{(N)}}+3{\Lambda^{(N-1)}}+{\Lambda^{(2)}}+
{\Lambda^{(1)}}$& 1&
$R_{20}$ &${\Lambda^{(N)}}+3{\Lambda^{(N-1)}}+{\Lambda^{(2)}}+
{\Lambda^{(1)}}$&-1\\
$R_{21}$&${\Lambda^{(N)}}+2{\Lambda^{(N-1)}}+{\Lambda^{(N-2)}}+
2{\Lambda^{(2)}}$& -1&
$R_{22}$&$2{\Lambda^{(N)}}+2{\Lambda^{(N-1)}}+2{\Lambda^{(1)}}$& 1\\
$R_{23}$&$ 2{\Lambda^{(N)}}+2{\Lambda^{(N-1)}}+2{\Lambda^{(1)}}$& -1&
$R_{24}$&$ 2{\Lambda^{(N)}}+{\Lambda^{(N-1)}}+{\Lambda^{(N-2)}}+
{\Lambda^{(2)}}+{\Lambda^{(1)}}$& 1\\
$R_{25}$&$3{\Lambda^{(N)}}+{\Lambda^{(N-2)}}+2{\Lambda^{(1)}}$& -1&
$R_{26}$&$ 3{\Lambda^{(N)}}+{\Lambda^{(N-1)}}+{\Lambda^{(1)}}$& -1
\\ \hline
\end{tabular}
}
Using the above data, the composite invariants can be computed. 
The composite polynomials for some of the framed $p$ knots are 
{\small
\begin{eqnarray}
{\cal H}_{(\twohor,\twohor)}[{\cal K}_3]&=&(dim_q \rho_{04}){q^{-4}}{\lambda^{2p}} \left( {\lambda^{4}}(1+ 2 q^3+2q^4+3 q^{6}+2 q^{7}+q^{8}+2 q^{9}+2 q^{10}+q^{12}
)\right.\nonumber \\
&& + {\lambda^{5}}(-2 q 
- 2 q^{2}+q^{3}-4 q^{4}-6 q^{5} -3 q^{7}-6 q^{8}-q^{9}-2 q^{10}-4 q^{11}-q^{13}-2
 q^{14})\nonumber \\
&& + {\lambda^{6}}(q^{2}+4 q^{3}-q^{4}+7 q^{6}+2 q^{7}
-3 q^{8}+6 q^{9}+4 q^{10}
- 2 q^{11}+3 q^{12}+2 q^{13}\nonumber \\
&& -2 q^{14}+2 q^{15}+q^{16})
+ {\lambda^{7}}(-2 q^{4}-q^{5}+4 q^{6}-3 q^{7}-6 q^{8}
+6 q^{9}-8 q^{11}+4 q^{12}\nonumber \\
&& + 2 q^{13}-6 q^{14}+q^{15}+2 q^{16}-q^{27})
+ {\lambda^{8}}(q^{6}-2 q^{7}-q^{8}+6 q^{9}-3 q^{10}-6 q^{11}\nonumber \\
&& +9q^{12}-8 q^{14}
+ 4 q^{15}+4 q^{16}-2q^{17}-q^{18}   )
+ {\lambda^{9}}(q^{9}-2 q^{10}-q^{11}+6 q^{12}-3 q^{13}\nonumber \\
&& -6q^{14} +6 q^{15}+2 q^{16}
-4 q^{17}
+ q^{19} )\nonumber \\
&& \left. + {\lambda^{10}}(+q^{12}-2 q^{13}-q^{14}+4q^{15}-q^{16}-2 q^{17}
+q^{18}  )\right)
\end{eqnarray}
\begin{eqnarray}
{\cal H}_{(\twohor,\twohor)}[{\cal K}_5]&=&(dim_q \rho_{04}){q^{-8}}{\lambda^{2p}}\left( {\lambda^{8}}(1+2 q^3+2q^4+3q^{6}+4q^{7}+3q^{8}+4q^{9}+ 6q^{10}
+4q^{11}+7q^{12}\right. \nonumber\\
&&+6q^{13} +5q^{14}+6q^{15}+7q^{16}+4 q^{17}+ 5q^{18}
+4q^{19}+3q^{20}+2q^{21}+2q^{22}+q^{24})\nonumber \\
&& + {\lambda^{9}}(-2 q-2q^{2}+q^{3} -4 q^{4}-8 q^{5}-2 q^{6}-5 q^{7}-14 q^{8}
- 9 q^{9}-10q^{10}-17 q^{11}\nonumber \\
&& - 12 q^{12}-15q^{13}-20 q^{14}-12q^{15}-14q^{16}-19 q^{17}-10 q^{18}
- 9q^{19}-12 q^{20}-6 q^{21}\nonumber \\
&& -4 q^{22}-6 q^{23}-2q^{24}-q^{25}
- 2q^{26} )
+ {\lambda^{10}}(q^{2}+4 q^{3}-q^{4}+11 q^{6}+6 q^{7}-q^{8}\nonumber \\
&&
+ 16 q^{9}+ 13q^{10}+6 q^{11}+18q^{12}+16 q^{13}+10q^{14}+24q^{15}
+15q^{16}+8q^{17}+21q^{18}\nonumber \\
&& +14 q^{19}+2q^{20}
+ 14 q^{21}+8 q^{22}+6q^{24}+4q^{25}-2q^{26}+2q^{27}+q^{28})\nonumber \\
&&
+ {\lambda^{11}}(-2 q^{4}-q^{5}+4 q^{6}-5 q^{7}-8 q^{8}
+6 q^{9}-4 q^{10}
- 14 q^{11}+2q^{12}-5 q^{13}-16 q^{14}\nonumber \\
&& - q^{15}-8 q^{16}-16 q^{17}+2q^{18}-6 q^{19}
-16 q^{20}+4q^{21}
- 2q^{22}-12q^{23}+4q^{24}+q^{25}\nonumber \\
&& -6q^{26}+2q^{27}+2q^{28}-q^{29})
+ {\lambda^{12}}(q^{6}-2q^{7}-q^{8}+6q^{9}-3q^{10}-6q^{11}
+11q^{12}\nonumber \\
&& -9q^{14}
+ 12q^{15}+3q^{16}-10q^{17}+12q^{18}+2q^{19}- 9q^{20}+12q^{21}+2q^{22}
-12 q^{23}\nonumber \\
&& + 10q^{24}
+ 2 q^{25}-10q^{26}+4q^{27}+3q^{28}-2q^{29})
+ {\lambda^{13}}(q^{9}-2q^{10}-q^{11}+6 q^{12}\nonumber \\
&& -4 q^{13}-6 q^{14}+11q^{15}-2q^{16}
- 11q^{17}+12 q^{18}-12q^{20}+11q^{21}-11q^{23}\nonumber \\
&& +12 q^{24}+q^{25}-12q^{26}+7q^{27}
+ 4q^{28}-4q^{29} )
+ {\lambda^{14}}(+q^{12}-2 q^{13}-q^{14}+6q^{15}\nonumber \\
&& -4q^{16}-6q^{17}+11 q^{18}-2q^{19}
- 11q^{20}+12q^{21}-12q^{23}+11 q^{24}-11q^{26}+8q^{27}\nonumber \\
&& +4q^{28}-4q^{29}
+q^{30}- q^{32})
+ {\lambda^{15}}(q^{15}-2 q^{16}-q^{17}+6q^{18}-4 q^{19}-6 q^{20}
+11 q^{21}\nonumber \\
&& - 2q^{22}-10 q^{23}+12 q^{24}-12 q^{26}+6 q^{27}+4 q^{28}-3 q^{29}
- q^{31}+q^{33} )\nonumber \\
&&
+{\lambda^{16}}( q^{18}-2q^{19}-q^{20}+6q^{21}-4q^{22}
-6q^{23}+11q^{24}
-2 q^{25}-11 q^{26}+8q^{27}\nonumber \\
&& +5q^{28}-4q^{29}-2q^{30}
+q^{32} )
+ {\lambda^{17}}(q^{21}-2 q^{22}-q^{23}+6q^{24}-4 q^{25}-6q^{26}+8q^{27}\nonumber \\
&&\left.
+ 2q^{28}-5q^{29}+q^{31} )
+ {\lambda^{18}}(q^{24}-2 q^{25}-q^{26}+4q^{27}-q^{28}-2q^{29}+
q^{30})\right)
\end{eqnarray}
\begin{eqnarray}
{\cal H}_{(\twohor,\twohor)}[{\cal K}_7]&=&(dim_q \rho_{04})q^{-12} {\lambda^{2p}}\left({\lambda^{12}} ( 1+2q^3+ 2 q^4+3 q^{6}+4 q^{7}+3 q^{8}
+4 q^{9}+6 q^{10}+6 q^{11}+9 q^{12}\right.\nonumber\\
&&+ 8 q^{13}+9 q^{14}+12 q^{15}+13 q^{16} +10 q^{17}+15 q^{18}+14 q^{19}+ 13 q^{20}+14 q^{21}+15 q^{22}\nonumber \\
&& +12 q^{23}+14 q^{24}
+12 q^{25}+11 q^{26}+ 10 q^{27}+9q^{28}+6 q^{29}+7 q^{30}+4 q^{31}
+3 q^{32}\nonumber \\
&& +2 q^{33}+2 q^{34}+q^{36})
+ {\lambda^{13}}(-2 q-2 q^{2}+q^{3}-4 q^{4}-8 q^{5}
- 2q^{6}-5 q^{7}-14 q^{8}\nonumber \\
&& -11 q^{9}
-12q^{10}-19 q^{11}-20 q^{12}-25 q^{13}-30 q^{14}
-26 q^{15}
- 34q^{16}-43 q^{17}\nonumber \\
&& -34 q^{18}-38 q^{19}-48 q^{20}
-40 q^{21}-40q^{22}
- 46 q^{23}-38 q^{24}-36 q^{25}-38 q^{26}\nonumber \\
&& -30 q^{27}-26q^{28}-26 q^{29}
- 18 q^{30}-15 q^{31}-14 q^{32}-8 q^{33}-6q^{34}-6 q^{35}-2 q^{36}\nonumber \\
&&
- q^{37}-2 q^{38})
+{\lambda^{14}}( q^{2}+4 q^{3}-q^{4}+11 q^{6}+6 q^{7}-q^{8}
+ 16 q^{9}+17 q^{10}+10 q^{11}\nonumber \\
&& + 22 q^{12}+26 q^{13}+22 q^{14}
+40 q^{15}+33 q^{16}+30 q^{17}+54 q^{18}+48 q^{19}
+33 q^{20}\nonumber \\
&& 
+ 60 q^{21}+54 q^{22}+38 q^{23}+54 q^{24}+50 q^{25}
+32 q^{26}+46 q^{27}
+ 36 q^{28}+22 q^{29}\nonumber \\
&& +29 q^{30}+24 q^{31}
+9 q^{32}+16 q^{33}+10 q^{34}
+ 2 q^{35}+6 q^{36}+4 q^{37}-2 q^{38}+2 q^{39}\nonumber \\
&& +q^{40} )
+ {\lambda^{15}}(-2q^{4}-q^{5}+4 q^{6}-5 q^{7}-8 q^{8}+6q^{9}
-4q^{10}-16q^{11}-7q^{13}-20q^{14}\nonumber \\
&& -5 q^{15}-18q^{16}-28q^{17}-6 q^{18}-26q^{19}
- 40 q^{20}- 9 q^{21}-26 q^{22}-45 q^{23}-12 q^{24}\nonumber \\
&& -22 q^{25}-40 q^{26}-10 q^{27}-18q^{28}-30 q^{29}
- 4q^{30}-10q^{31}-20 q^{32}+q^{33}-2q^{34}\nonumber \\
&& -11 q^{35}
+4q^{36}
+ q^{37}-6 q^{38}+2q^{39}+2 q^{40}-q^{41} )
+ {\lambda^{16}}(q^{6}-2 q^{7}-q^{8}+6 q^{9}\nonumber \\
&& -3 q^{10}-6 q^{11}
+11q^{12}- 9 q^{14}+12 q^{15}+3 q^{16}-8 q^{17}+16 q^{18}+2 q^{19}
-7 q^{20}\nonumber \\
&& +22 q^{21}
+3q^{22}
- 12q^{23}+26 q^{24}+6q^{25}-14q^{26}+22q^{27}+9 q^{28}-14q^{29}\nonumber \\
&& +17q^{30}+4q^{31}
-12q^{32}+ 12q^{33}+2q^{34}-12q^{35}+ 9 q^{36}+2q^{37}-9q^{38}+4q^{39}\nonumber \\
&& +3q^{40}
- 2q^{41})
+ {\lambda^{17}}(q^{9}-2 q^{10}-q^{11}+6 q^{12}-4 q^{13}-6 q^{14}+11 q^{15}-2q^{16}\nonumber \\
&&
- 11 q^{17}+12 q^{18}-12 q^{20}+13 q^{21}-2 q^{22}-13 q^{23}+18 q^{24}- 4 q^{25}-18 q^{26}\nonumber \\
&&
+21 q^{27}-21 q^{29}+18 q^{30}+4 q^{31}-18q^{32}+14 q^{33}+2 q^{34}-14 q^{35}+12 q^{36}\nonumber \\
&&
+ q^{37}-12q^{38}+6 q^{39} +4 q^{40}-3 q^{41} )
+ {\lambda^{18}}(+q^{12}-2 q^{13}-q^{14}+6 q^{15}-4 q^{16}\nonumber \\
&& -6 q^{17}
+11 q^{18}-2 q^{19}-11 q^{20}
+ 12 q^{21}-12 q^{23} +13 q^{24}-2 q^{25}-13 q^{26}\nonumber \\
&& +18 q^{27}-4 q^{28}
-18 q^{29}+21 q^{30}
- 21 q^{32}+18 q^{33}+4 q^{34}-18q^{35}+14 q^{36}\nonumber \\
&& +2 q^{37}-14q^{38}+8 q^{39}
+ 4 q^{40}-4 q^{41} )
+ {\lambda^{19}}(q^{15}-2 q^{16}-q^{17}+6 q^{18}-4 q^{19}\nonumber \\
&& -6 q^{20}
+ 11 q^{21}-2 q^{22}
- 11q^{23}+12q^{24}-12q^{26}+13q^{27}-2 q^{28}-13q^{29}\nonumber \\
&& +18q^{30}-4 q^{31}-
18 q^{32}
+ 21 q^{33}-21 q^{35}+18 q^{36}+4 q^{37}-18 q^{38}+9 q^{39}\nonumber \\
&& +6 q^{40}-4
    q^{41}-q^{43} )
+{\lambda^{20}}( q^{18}-2 q^{19}-q^{20}+6 q^{21}-4q^{22}-6 q^{23}
+11 q^{24}\nonumber \\
&& -2 q^{25}
- 11 q^{26}+12 q^{27}-12 q^{29}+13 q^{30}-2 q^{31}- 13 q^{32}
+18 q^{33}-4 q^{34}\nonumber \\
&& -18 q^{35}
+ 21 q^{36}-21 q^{38}+12 q^{39}+9 q^{40}-6 q^{41}-q^{42}+q^{44}-q^{46})\nonumber \\
&&
+{\lambda^{21}}( q^{21}-2 q^{22}-q^{23}+6 q^{24}-4 q^{25}-6 q^{26}+11 q^{27}-2 q^{28}-11 q^{29}
+ 12 q^{30}\nonumber \\
&& -12 q^{32}+14 q^{33}-2 q^{34}-13 q^{35}+18 q^{36}-4 q^{37}-18 q^{38}
+ 14 q^{39}+6 q^{40}-9 q^{41}\nonumber \\
&& +2 q^{43}-q^{45}+q^{47})
+{\lambda^{22}} (q^{24}-2 q^{25}-q^{26}+6 q^{27}-4 q^{28}-6 q^{29}+11 q^{30}\nonumber \\
&& -2 q^{31}
- 11 q^{32}+12 q^{33}-12q^{35}+13 q^{36}-2 q^{37}-13 q^{38}+12 q^{39}+3 q^{40}-6q^{41}\nonumber \\
&&
+ 2 q^{42}-2 q^{44}+q^{46})
+ {\lambda^{23}}(q^{27}-2 q^{28}-q^{29}+6 q^{30}-4 q^{31}-6 q^{32}+11 q^{33}\nonumber \\
&& -2 q^{34}
- 11 q^{35}+12 q^{36}-12 q^{38}+8 q^{39}+4 q^{40}-3q^{41}-2q^{43}+q^{45})\nonumber \\
&&
+ {\lambda^{24}}(q^{30}-2 q^{31}-q^{32}+6 q^{33}-4 q^{34}
-6 q^{35}+11q^{36}-2 q^{37}
- 11 q^{38}+8 q^{39}\nonumber \\
&& +5 q^{40}-4 q^{41}-2 q^{42}+q^{44})
+ {\lambda^{25}}(q^{33}-2 q^{34}-q^{35}+6 q^{36}-4q^{37}-6 q^{38} \\
&&\left. +8 q^{39} +2 q^{40}-5 q^{41}+q^{43} )
+ {\lambda^{26}}(q^{36}-2 q^{37}-q^{38}+4 q^{39}-q^{40}
-2 q^{41} +q^{42} )\right)\nonumber 
\end{eqnarray}
}
\subsection{(R,S)=(two vertical box, two vertical box)}
Let us denote the composite representation $(\twover,\twover)$ by $\rho_{05}$
and its highest weight is
$\rho_{05} = {\Lambda^{(N-2)}}+{\Lambda^{(2)}}$.
The representations $R_t$ 
obtained from $\rho_{05}\otimes\rho_{05}$ and the signs of the 
braiding eigenvalues $\epsilon_t$ are
{\small
\begin{displaymath}
\begin{array}{ll}
R_1 = {\Lambda^{(N-2)}}+{\Lambda^{(N-2)}}+{\Lambda^{(2)}}+{\Lambda^{(2)}};\epsilon_{1} = 1\qquad&
R_2 = {\Lambda^{(N-1)}}+{\Lambda^{(N-3)}}+{\Lambda^{(2)}}+{\Lambda^{(2)}};\epsilon_{2} = -1\\
R_3 = {\Lambda^{(N)}}+{\Lambda^{(N-4)}}+{\Lambda^{(2)}}+{\Lambda^{(2)}};\epsilon_{3} = 1&
R_4 = {\Lambda^{(N-1)}}+{\Lambda^{(N-2)}}+{\Lambda^{(2)}}+{\Lambda^{(1)}};\epsilon_{4} = 1\\
R_5 = {\Lambda^{(N)}}+{\Lambda^{(N-3)}}+{\Lambda^{(2)}}+{\Lambda^{(1)}};\epsilon_{5} = 1&
R_6 = {\Lambda^{(N)}}+{\Lambda^{(N-2)}}+{\Lambda^{(1)}}+{\Lambda^{(1)}};\epsilon_{6} = -1\\
R_7 = {\Lambda^{(N-2)}}+{\Lambda^{(N-2)}}+{\Lambda^{(3)}}+{\Lambda^{(1)}};\epsilon_{7} = -1&
R_8 = {\Lambda^{(N-1)}}+{\Lambda^{(N-3)}}+{\Lambda^{(3)}}+{\Lambda^{(1)}};\epsilon_{8} = 1\\
R_9 = {\Lambda^{(N)}}+{\Lambda^{(N-4)}}+{\Lambda^{(3)}}+{\Lambda^{(1)}};\epsilon_{9} = -1&
R_{10} = {\Lambda^{(N-1)}}+{\Lambda^{(N-1)}}+{\Lambda^{(2)}};
\epsilon_{10} = - 1\\
R_{11} = {\Lambda^{(N)}}+{\Lambda^{(N-2)}}+{\Lambda^{(2)}};\epsilon_{11} = 1&
R_{12} = {\Lambda^{(N-1)}}+{\Lambda^{(N-1)}}+{\Lambda^{(1)}}+{\Lambda^{(1)}};\epsilon_{12} = 1\\
R_{13} = {\Lambda^{(N-1)}}+{\Lambda^{(N-2)}}+{\Lambda^{(3)}};\epsilon_{13} = 1&
R_{14} = {\Lambda^{(N)}}+{\Lambda^{(N-3)}}+{\Lambda^{(3)}};\epsilon_{14} = 1\\
R_{15} = {\Lambda^{(N-2)}}+{\Lambda^{(N-2)}}+{\Lambda^{(4)}};\epsilon_{15} = 1&
R_{16} = {\Lambda^{(N-1)}}+{\Lambda^{(N-3)}}+{\Lambda^{(4)}};\epsilon_{16} = -1\\
R_{17} = {\Lambda^{(N)}}+{\Lambda^{(N-4)}}+{\Lambda^{(4)}};\epsilon_{17} = 1&
R_{18} = {\Lambda^{(N)}}+{\Lambda^{(N-1)}}+{\Lambda^{(1)}};\epsilon_{18} = 1\\
R_{19} = 2{\Lambda^{(N)}};\epsilon_{19} = 1&
R_{20} = {\Lambda^{(N-1)}}+{\Lambda^{(N-2)}}+{\Lambda^{(2)}}+{\Lambda^{(1)}};\epsilon_{20} = -1\\
R_{21} = {\Lambda^{(N)}}+{\Lambda^{(N-3)}}+{\Lambda^{(2)}}+{\Lambda^{(1)}};\epsilon_{21} = -1&
R_{22} = {\Lambda^{(N)}}+{\Lambda^{(N-2)}}+{\Lambda^{(2)}};\epsilon_{22} = -1\\
R_{23} = {\Lambda^{(N)}}+{\Lambda^{(N-2)}}+{\Lambda^{(2)}};\epsilon_{23} = 1&
R_{24} = {\Lambda^{(N-1)}}+{\Lambda^{(N-2)}}+{\Lambda^{(3)}};\epsilon_{24} = -1\\
R_{25} = {\Lambda^{(N)}}+{\Lambda^{(N-3)}}+{\Lambda^{(3)}};\epsilon_{25} =-1&
R_{26} = {\Lambda^{(N)}}+{\Lambda^{(N-1)}}+{\Lambda^{(1)}};\epsilon_{26} = -1
\end{array}
\end{displaymath}
}
The composite invariants of framed knots and links can be computed 
using the above data. Some of the framed $p$ knot polynomials are
{\small
\begin{eqnarray}
{\cal H}_{(\twover,\twover)}[{\cal K}_3]&=&(dim_q \rho_{05}){q^{-15}}{\lambda^{2p}}\left( {\lambda^{4}}(q^{7}+2 q^{9}+2 q^{10}+q^{11}+2q^{12}+3 q^{13}+2 q^{15}+2 q^{16}+q^{19} )\right.\nonumber \\
&&
+ {\lambda^{5}}(- 2q^{5}- q^{6}-4 q^{8}-2 q^{9}-q^{10}-6 q^{11}-3 q^{12}-6 q^{14}-4
q^{15}+q^{16}- 2 q^{17}-2 q^{18})\nonumber \\
&&
+ {\lambda^{6}}(q^{3}+2 q^{4}-2 q^{5}+2 q^{6}+3 q^{7}-2 q^{8}
+4 q^{9}+6 q^{10}-3 q^{11}+2 q^{12}+ 7 q^{13}
- q^{15}\nonumber \\
&& +4 q^{16}+q^{17})
+ {\lambda^{7}}(-q^{2}+2 q^{3}+q^{4}-6 q^{5}+2q^{6}+4 q^{7}
-8 q^{8}+6 q^{10}- 6 q^{11}-3 q^{12}\nonumber \\
&&
+ 4 q^{13}-q^{14}-2 q^{15})
+ {\lambda^{8}}(-q-2 q^{2}+4 q^{3}+4 q^{4}-8 q^{5}+9 q^{7}
-6 q^{8}-3 q^{9}+ 6 q^{10}\nonumber \\
&&
- q^{11}-2 q^{12}+q^{13})
+ {\lambda^{9}}(1-4 q^{2}+2 q^{3}+6q^{4}-6 q^{5}-3q^{6}+6 q^{7}
-q^{8}-2 q^{9}+q^{10} )\nonumber \\
&&\left.
+ {\lambda^{10}}(q-2 q^{2}-q^{3}+4 q^{4}-q^{5}-2 q^{6}+q^{7})\right)
\end{eqnarray}
\begin{eqnarray}
{\cal H}_{(\twover,\twover)}[{\cal K}_5]&=&(dim_q \rho_{05}){q^{-25}}{\lambda^{2p}}\left( {\lambda^{8}}(q^{-16}+2q^{-14}+2q^{-13}+ 3q^{-12}+4q^{-11}
+5q^{-10}+ 4q^{-9}+7q^{-8}\right.\nonumber\\
&&+6q^{-7} +5q^{-6}+ 6 q^{-5}+ 7q^{-4}+4 q^{-3}+ 6 q^{-2}
+4q^{-1}+3+4q+3q^{2}+ 2q^{4}+2q^{5}+q^{8})\nonumber \\
&&
+ {\lambda^{9}}(- 2q^{-18}-q^{-17}-2q^{-16}-6 q^{-15}
- 4 q^{-14}-6q^{-13}-12q^{-12}-9 q^ {-11}- 10 q^{-10}\nonumber \\
&&
- 19 q^{-9}- 14 q^{-8}-12 q^{-7}-20q^{-6}-15 q^{-5}
- 12 q^{-4}- 17 q^{-3}-10 q^{-2}-9 q^{-1}\nonumber \\
&& -14
- 5 q-2 q^{2}- 8 q^{3}-4 q^{4}+q^{5}- 2 q^{6}-2q^{7})
+ {\lambda^{10}}(q^{-20}+ 2 q^{-19}-2 q^{-18}\nonumber \\
&& +4q^{-17}
+ 6 q^{-16}+8 q^{-14}+14 q^{-13}+ 2 q^{-12}+14 q^{-11}+21 q^{-10}+8 q^{-9}
+ 15 q^{-8}\nonumber \\
&& +24 q^{-7}+ 10 q^{-6}+16 q^{-5}+ 18 q^{-4}+6 q^{-3}+ 13 q^{-2}+16 q^{-1}
-1+ 6q+11 q^{2}\nonumber \\
&& -q^{4}+4 q^{5}+q^{6})
+ {\lambda^{11}}(-q^{-21}+ 2q^{-20}+2 q^{-19}-6 q^{-18}+ q^{-17}+4 q^{-16}- 12q^{-15}\nonumber \\
&& - 2q^{-14}
+4 q^{-13}- 16 q^{-12}-6 q^{-11}+2 q^{-10}-16 q^{-9}-8 q^{-8}-q^{-7}- 16 q^{-6}- 5 q^{-5}\nonumber \\
&&
+2 q^{-4}-14 q^{-3}-4 q^{-2}+6 q^{-1}-8 - 5 q
+4 q^{2}- q^{3}- 2 q^{4})
- {\lambda^{12}} (2 q^{-21}+3 q^{-20}\nonumber \\
&& + 4 q^{-19}-10 q^{-18}+2q^{-17}+ 10 q^{-16}-12 q^{-15}
+  2 q^{-14}+ 12q^{-13}-9 q^{-12}+ 2q^{-11}\nonumber \\
&& +12 q^{-10}-10 q^{-9}+ 3 q^{-8}+ 12 q^{-7}-9 q^{-6}+11 q^{-4}-6 q^{-3}
- 3 q^{-2}+6 q^{-1}-1\nonumber \\
&& - 2 q+ q^{2})
+ {\lambda^{13}}(-4 q^{-21}+4 q^{-20}+ 7q^{-19}-12 q^{-18}+q^{-17}+ 12 q^{-16}-11 q^{-15}\nonumber \\
&& +11 q^{-13}- 12 q^{-12}
+ 12 q^{-10}- 11 q^{-9}-2 q^{-8}+ 11 q^{-7}-6 q^{-6}- 4q^{-5}+6 q^{-4}\nonumber \\
&& -q^{-3}-2 q^{-2}+q^{-1}
)
+ {\lambda^{14}}(-q^{-24}+q^{-22}- 4 q^{-21}+4 q^{-20}+8 q^{-19}-11 q^{-18}\nonumber \\
&& +11 q^{-16}
- 12 q^{-15}+12 q^{-13}-11 q^{-12}-2 q^{-11}+11 q^{-10}-6 q^{-9}- 4 q^{-8}+6 q^{-7}\nonumber \\
&&
- q^{-6}-2 q^{-5}
+ q^{-4})
+{\lambda^{15}}( q^{-25}-q^{-23}-3 q^{-21}+4 q^{-20}+6 q^{-19}-12 q^{-18}\nonumber \\
&& +12 q^{-16}-10 q^{-15}- 2 q^{-14}+11q^{-13}
- 6 q^{-12}-4 q^{-11}+6 q^{-10}-q^{-9}-2q^{-8}+q^{-7})\nonumber \\
&&
+ {\lambda^{16}}(q^{-24}-2 q^{-22}-4 q^{-21}+5 q^{-20}+ 8 q^{-19}-11 q^{-18}-2 q^{-17}+ 11 q^{-16}-6 q^{-15}\nonumber \\
&&
- 4 q^{-14}+6 q^{-13}-q^{-12}-2 q^{-11}+q^{-10})
+ {\lambda^{17}}(q^{-23}- 5 q^{-21}+2 q^{-20}+8 q^{-19}\nonumber \\
&& -6 q^{-18}-4 q^{-17}+6q^{-16}- q^{-15}-2 q^{-14}+q^{-13})
+ {\lambda^{18}}(q^{-22}-2 q^{-21}-q^{-20}\nonumber \\
&&\left. + 4 q^{-19}-q^{-18}-2 q^{-17}+q^{-16})\right)
\end{eqnarray}
}
These composite invariants ${\cal H}_{(R,R)}[{\cal K}]$
and the corresponding $SO(N)$ invariants ${\cal G}_R[{\cal K}]$
given in appendix A of Ref.\cite {prav} satisfy the
generalised Rudolph theorem (\ref {rud}).
\section{$SO(N)$ Reformulated Invariants $g_{R_1,R_2,\ldots,R_r}(q,\lambda)$}
Here we present the $SO$ reformulated invariants for framed
knots and links.
\subsection{$p$ Framed Torus Knot $(2,5)$}
\begin{equation}
g_{\one}={\left( -1 \right) }^p\,q^{-1}\,{\lambda^{{\p2}+1}} \,\left( 1 - q^2\,\left( -1 + \lambda  \right)  - \lambda  + 
      q\,\left( 1 - \lambda  + {\lambda }^2 \right)  \right) 
\end{equation}
\begin{eqnarray}
g_{\twohor} &=& \frac{-{\lambda^p}}{2\,{\left( -1 + q \right) }^2\,q^4} \left( \left( -1 + \lambda  \right) \,\lambda \,
      \left( 2\,q^{\frac{5}{2} + p}\,{\sqrt{\lambda }} + 2\,q^{\frac{9}{2} + p}\,{\sqrt{\lambda }} - 
        2\,q^{\frac{19}{2} + p}\,{\sqrt{\lambda }} \right.\right.\nonumber \\
&& + 2\,q^{\frac{5}{2}}\,\left( -1 + \lambda  \right) \,{\sqrt{\lambda }} - 
        2\,q^{\frac{15}{2}}\,\left( -1 + \lambda  \right) \,{\sqrt{\lambda }} + 
        2\,q^{\frac{7}{2}}\,\left( 1 - 2\,\lambda  \right) \,{\lambda }^{\frac{3}{2}} \nonumber \\
&& + q\,\left( -1 + \lambda  \right) \,{\lambda }^2 + q^9\,\left( -1 + \lambda  \right) \,{\lambda }^2 + 
        2\,q^{\frac{11}{2} + p}\,{\lambda }^{\frac{5}{2}} - 2\,q^{\frac{25}{2} + p}\,{\lambda }^{\frac{5}{2}} + 
        2\,q^{9 + p}\,{\lambda }^3 \nonumber \\
&& + 2\,q^{13 + p}\,{\lambda }^3 - \left( -1 + \lambda  \right) \,{\lambda }^3 - 
        q^{10}\,\left( -1 + \lambda  \right) \,{\lambda }^3 - 
        2\,q^{\frac{7}{2} + p}\,{\sqrt{\lambda }}\,\left( 1 + \lambda  \right) \nonumber \\
&& - 2\,q^{\frac{13}{2} + p}\,{\lambda }^{\frac{3}{2}}\,\left( 1 + \lambda  \right)  + 
        2\,q^{\frac{15}{2} + p}\,{\lambda }^{\frac{3}{2}}\,\left( 1 + \lambda  \right)  - 
        2\,q^{\frac{21}{2} + p}\,{\lambda }^{\frac{5}{2}}\,\left( 1 + \lambda  \right) \nonumber \\
&& -  2\,q^{10 + p}\,{\lambda }^2\,{\left( 1 + \lambda  \right) }^2 - 
        2\,q^{12 + p}\,{\lambda }^2\,{\left( 1 + \lambda  \right) }^2 + 
        2\,q^{\frac{13}{2}}\,{\lambda }^{\frac{3}{2}}\,\left( -1 + 2\,\lambda  \right) \nonumber \\
&& + 
        2\,q^{\frac{23}{2} + p}\,{\lambda }^{\frac{3}{2}}\,\left( 1 + \lambda  + {\lambda }^2 \right)  + 
        4\,q^{11 + p}\,{\lambda }^2\,\left( 1 + \lambda  + {\lambda }^2 \right)  \nonumber \\
&& + 
        q^2\,\lambda \,\left( -1 + \lambda  + 2\,{\lambda }^2 - 2\,{\lambda }^3 \right)  + 
        q^8\,\lambda \,\left( -1 + \lambda  + 2\,{\lambda }^2 - 2\,{\lambda }^3 \right) \nonumber \\
&& + 
        2\,q^{\frac{9}{2}}\,{\sqrt{\lambda }}\,\left( -1 + \lambda  + {\lambda }^2 + {\lambda }^3 \right)  - 
        2\,q^{\frac{11}{2}}\,{\sqrt{\lambda }}\,\left( -1 + \lambda  + {\lambda }^2 + {\lambda }^3 \right) \nonumber \\
&& + 
        q^3\,\left( 1 - \lambda  - 2\,{\lambda }^2 - {\lambda }^4 + {\lambda }^5 \right)  + 
        q^7\,\left( 1 - \lambda  - 2\,{\lambda }^2 - {\lambda }^4 + {\lambda }^5 \right) \nonumber \\
&& + 
        q^4\,\lambda \,\left( -2 + 2\,\lambda  + 6\,{\lambda }^2 - 2\,{\lambda }^3 + {\lambda }^4 + {\lambda }^5 + 
           {\lambda }^6 + {\lambda }^7 \right) \nonumber \\
&&  + q^6\,\lambda \,
         \left( -2 + 2\,\lambda  + 6\,{\lambda }^2 - 2\,{\lambda }^3 + {\lambda }^4 + {\lambda }^5 + {\lambda }^6 + 
           {\lambda }^7 \right) \nonumber \\
&& \left.\left. - 2\,q^5\,\left( -1 + 2\,{\lambda }^2 + 2\,{\lambda }^4 + {\lambda }^6 + {\lambda }^7 + 
           {\lambda }^8 \right)  \right)  \right)
\end{eqnarray}
\begin{eqnarray}
g_{\twover} &=& \frac{-{\lambda^p}}{2\,{\left( -1 + q \right) }^2} \left( q^{-7 - p}\,\left( -1 + \lambda  \right) \,\lambda \,
      \left( 2\,q^{\frac{7}{2}}\,{\sqrt{\lambda }} - 2\,q^{\frac{17}{2}}\,{\sqrt{\lambda }} - 
        2\,q^{\frac{21}{2}}\,{\sqrt{\lambda }} \right.\right.\nonumber \\
&& + 2\,q^{\frac{11}{2} + p}\,\left( -1 + \lambda  \right) \,{\sqrt{\lambda }} - 
        2\,q^{\frac{21}{2} + p}\,\left( -1 + \lambda  \right) \,{\sqrt{\lambda }} + 
        q^{4 + p}\,\left( -1 + \lambda  \right) \,{\lambda }^2 \nonumber \\
&& + q^{12 + p}\,\left( -1 + \lambda  \right) \,{\lambda }^2 + 
        2\,{\sqrt{q}}\,{\lambda }^{\frac{5}{2}} - 2\,q^{\frac{15}{2}}\,{\lambda }^{\frac{5}{2}} + 2\,{\lambda }^3 + 
        2\,q^4\,{\lambda }^3 \nonumber \\
&& - q^{3 + p}\,\left( -1 + \lambda  \right) \,{\lambda }^3 - 
        q^{13 + p}\,\left( -1 + \lambda  \right) \,{\lambda }^3 + 
        2\,q^{\frac{19}{2}}\,{\sqrt{\lambda }}\,\left( 1 + \lambda  \right)\nonumber \\
&&  - 
        2\,q^{\frac{11}{2}}\,{\lambda }^{\frac{3}{2}}\,\left( 1 + \lambda  \right)  + 
        2\,q^{\frac{13}{2}}\,{\lambda }^{\frac{3}{2}}\,\left( 1 + \lambda  \right)  + 
        2\,q^{\frac{5}{2}}\,{\lambda }^{\frac{5}{2}}\,\left( 1 + \lambda  \right)  - 
        2\,q\,{\lambda }^2\,{\left( 1 + \lambda  \right) }^2 \nonumber \\
&& - 2\,q^3\,{\lambda }^2\,{\left( 1 + \lambda  \right) }^2 - 
        2\,q^{\frac{13}{2} + p}\,{\lambda }^{\frac{3}{2}}\,\left( -1 + 2\,\lambda  \right)  + 
        2\,q^{\frac{19}{2} + p}\,{\lambda }^{\frac{3}{2}}\,\left( -1 + 2\,\lambda  \right) \nonumber \\
&& - 
        2\,q^{\frac{3}{2}}\,{\lambda }^{\frac{3}{2}}\,\left( 1 + \lambda  + {\lambda }^2 \right)  + 
        4\,q^2\,{\lambda }^2\,\left( 1 + \lambda  + {\lambda }^2 \right)\nonumber \\
&&  + 
        q^{5 + p}\,\lambda \,\left( -1 + \lambda  + 2\,{\lambda }^2 - 2\,{\lambda }^3 \right)  + 
        q^{11 + p}\,\lambda \,\left( -1 + \lambda  + 2\,{\lambda }^2 - 2\,{\lambda }^3 \right) \nonumber \\
&& + 
        2\,q^{\frac{15}{2} + p}\,{\sqrt{\lambda }}\,\left( -1 + \lambda  + {\lambda }^2 + {\lambda }^3 \right)  - 
        2\,q^{\frac{17}{2} + p}\,{\sqrt{\lambda }}\,\left( -1 + \lambda  + {\lambda }^2 + {\lambda }^3 \right) \nonumber \\
&& + 
        q^{6 + p}\,\left( 1 - \lambda  - 2\,{\lambda }^2 - {\lambda }^4 + {\lambda }^5 \right)  + 
        q^{10 + p}\,\left( 1 - \lambda  - 2\,{\lambda }^2 - {\lambda }^4 + {\lambda }^5 \right) \nonumber \\
&& + 
        q^{7 + p}\,\lambda \,\left( -2 + 2\,\lambda  + 6\,{\lambda }^2 - 2\,{\lambda }^3 + {\lambda }^4 + {\lambda }^5 + 
           {\lambda }^6 + {\lambda }^7 \right) \nonumber \\
&& + q^{9 + p}\,\lambda \,
         \left( -2 + 2\,\lambda  + 6\,{\lambda }^2 - 2\,{\lambda }^3 + {\lambda }^4 + {\lambda }^5 + {\lambda }^6 + 
           {\lambda }^7 \right) \nonumber \\
&& \left.\left. - 2\,q^{8 + p}\,\left( -1 + 2\,{\lambda }^2 + 2\,{\lambda }^4 + {\lambda }^6 + 
           {\lambda }^7 + {\lambda }^8 \right)  \right)  \right)
\end{eqnarray}
\subsection{Connected sum of trefoil and trefoil with framing $p$}
The composite knot invariants for the 
connected sum of two knots ${\cal K}_1$ and
${\cal K}_2$ will be 
\begin{equation}
{\cal H}_{(R,S)}[{\cal K}_1\# {\cal K}_2]={ 
{\cal H}_{(R,S)}[{\cal K}_1] {\cal H}_{(R,S)}[{\cal K}_2]\over 
{\cal H}_{(R,S)}[U]}~.
\end{equation}
Similarly, the  $SO(N)$ invariants ${\cal G}_R[{\cal K}_1\# {\cal K}_2$ 
will also be product of $SO(N)$ invariants of the two knots 
normalised by the $SO(N)$ unknot invariant. From these
invariants, we can compute the $SO$ reformulated invariants
for the connected sum of two knots. For the connected sum of trefoil, the
reformulated polynomial for various representations are: 
\begin{eqnarray}
g_{\one} &=& {\left( -1 \right) }^p\,q^{-2}\,{\lambda }^{2+{\p2}}\,\left( {\left( -1 + \lambda  \right) }^2 + q^4\,{\left( -1 + \lambda  \right) }^2 + 
      {\sqrt{q}}\,\left( -1 + \lambda  \right) \,{\lambda }^{\frac{3}{2}}\right.\nonumber \\
&& - 
      q^{\frac{7}{2}}\,\left( -1 + \lambda  \right) \,{\lambda }^{\frac{3}{2}} + q^2\,\left( 2 - 4\,\lambda  + 3\,{\lambda }^2 \right)  - 
      q^{\frac{3}{2}}\,{\sqrt{\lambda }}\,\left( -1 + {\lambda }^3 \right) \nonumber \\
&& + 
      q^{\frac{5}{2}}\,{\sqrt{\lambda }}\,\left( -1 + {\lambda }^3 \right)  - 
      q\,\left( -2 + 3\,\lambda  - 2\,{\lambda }^2 + {\lambda }^3 \right)\nonumber \\
&& \left.  - q^3\,\left( -2 + 3\,\lambda  - 2\,{\lambda }^2 + {\lambda }^3 \right) 
      \right) 
\end{eqnarray}
\begin{eqnarray}
g_{\twohor}&=& \frac{{\lambda }^{3+p}}{2\,{\left( -1 + q \right) }^2\,q^4\,\left( 1 + q \right) }
\left[ \left( -1 - q \right) \right.\nonumber \\
&& {\left( -1 + \left( {\sqrt{q}} - {\sqrt{\lambda }} \right) 
        \left( {\sqrt{q}} \left( -q + q \lambda  - \lambda  \right)  - {\sqrt{\lambda }} \right)  \right) }^4
   {\left( {\sqrt{q}} + {\sqrt{\lambda }} \right) }^2 {\left( -1 + {\sqrt{q}} {\sqrt{\lambda }} \right) }^2 \nonumber \\
&& + 
  2\,q\,\left( 1 + q \right) \,{\left( -1 + \lambda  \right) }^2\,{\left( 1 + q^2 - q\,\lambda  \right) }^4  - 
  2\,q^{1 + p}\,\left( -1 + \lambda  \right) \,\left( -1 + q\,\lambda  \right) \nonumber \\
&&   {\left( 1 + q^4 - q^3\,\left( -1 + \lambda  \right)  - q^6\,\left( -1 + \lambda  \right)  - q^2\,\lambda  + 
       q^5\,\left( -1 + \lambda  \right) \,\lambda  \right) }^2 \nonumber \\
&& + 
  \left( 1 + q \right) \,\left( q - \lambda  \right) \,\left( -1 + q\,\lambda  \right) \,
   \left( {\left( 1 + q^2 \right) }^2 - \left( 2 + q\,\left( -1 + 2\,q \right) \,\left( 1 + q^2 \right)  \right) \,\lambda \right.\nonumber \\
&& \left. + 
       {\left( 1 + \left( -1 + q \right) \,q \right) }^2\,{\lambda }^2 + {\left( -1 + q \right) }^2\,q\,{\lambda }^3 \right)^2 \nonumber \\
&& + 
  2\,q^{\frac{1}{2} + p}\,\left( -1 + \lambda  \right) \left( -{\sqrt{q}} - {\sqrt{\lambda }} + q^2\,{\sqrt{\lambda }} + 
     q^{\frac{3}{2}}\,\lambda  \right) \nonumber \\
&& \left( -1 - q^{\frac{3}{2}}\,{\sqrt{\lambda }} + q\,\lambda  + q^2\,\lambda  + 
       q^{\frac{5}{2}}\,{\lambda }^{\frac{3}{2}} - q^{\frac{17}{2}}\,{\lambda }^{\frac{3}{2}} + q^{\frac{11}{2}}\,{\lambda }^{\frac{5}{2}} + 
       q^{\frac{19}{2}}\,{\lambda }^{\frac{5}{2}} + q^7\,{\lambda }^3 \right. \nonumber \\
&& + q^{\frac{7}{2}}\,{\sqrt{\lambda }}\,\left( 1 + \lambda  \right)  - 
       q^9\,{\lambda }^2\,\left( 1 + \lambda  \right)  - q^{\frac{9}{2}}\,{\sqrt{\lambda }}\,{\left( 1 + \lambda  \right) }^2 + 
       q^{\frac{13}{2}}\,{\sqrt{\lambda }}\,{\left( 1 + \lambda  \right) }^2 \nonumber \\
&& + q^8\,\lambda \,{\left( 1 + \lambda  \right) }^2 + 
       q^5\,\lambda \,\left( 2 + \lambda  \right)  - q^{\frac{15}{2}}\,{\lambda }^{\frac{3}{2}}\,\left( 1 + 2\,\lambda  \right)  + 
       q^4\,\left( -1 + \lambda  + {\lambda }^2 \right) \nonumber \\
&& \left.\left. - q^3\,\left( 1 + \lambda  + {\lambda }^2 \right)  - 
       q^6\,\left( 1 + \lambda  + 3\,{\lambda }^2 + {\lambda }^3 \right)  \right)^2 \right]
\end{eqnarray}
\begin{eqnarray}
g_{\twover}&=& \frac{{\lambda }^{3+p}}{2\,{\left( -1 + q \right) }^2\,q^4\,\left( 1 + q \right) }
\left[ \left( -1 - q \right) \right. \nonumber \\
&& {\left( -1 + \left( {\sqrt{q}} - {\sqrt{\lambda }} \right) 
        \left( {\sqrt{q}} \left( -q + q \lambda - \lambda  \right)  - {\sqrt{\lambda }} \right)  \right) }^4
   {\left( {\sqrt{q}} + {\sqrt{\lambda }} \right) }^2 {\left( -1 + {\sqrt{q}}\,{\sqrt{\lambda }} \right) }^2 \nonumber \\
&& + 
  2\,q\,\left( 1 + q \right) \,{\left( -1 + \lambda  \right) }^2\,{\left( 1 + q^2 - q\,\lambda  \right) }^4 + 
  2\,q^{-3 - p}\,\left( q - \lambda  \right) \,\left( -1 + \lambda  \right) \nonumber \\ 
&&   {\left( 1 + q^2 + q^6 - q^3\,\left( -1 + \lambda  \right)  - \lambda  - q^4\,\lambda  + q\,\left( -1 + \lambda  \right) \,\lambda  \right) }^2 \nonumber \\
&& +
   \left( 1 + q \right) \,\left( q - \lambda  \right) \,\left( -1 + q\,\lambda  \right) \,
   \left( {\left( 1 + q^2 \right) }^2 - \left( 2 + q\,\left( -1 + 2\,q \right) \,\left( 1 + q^2 \right)  \right) \,\lambda \right.\nonumber \\
&& \left. + 
       {\left( 1 + \left( -1 + q \right) \,q \right) }^2\,{\lambda }^2 + {\left( -1 + q \right) }^2\,q\,{\lambda }^3 \right)^2 \nonumber \\
&& + 
  2\,q^{-\left( \frac{21}{2} \right)  - p}\,\left( -1 + \lambda  \right) \,
   \left( -q^{\frac{3}{2}} - {\sqrt{\lambda }} + q^2\,{\sqrt{\lambda }} + {\sqrt{q}}\,\lambda  \right) \nonumber \\
&&   \left( q^{\frac{19}{2}} - q^8\,{\sqrt{\lambda }} - q^{\frac{15}{2}}\,\lambda  - q^{\frac{17}{2}}\,\lambda  - q\,{\lambda }^{\frac{3}{2}} + 
       q^7\,{\lambda }^{\frac{3}{2}} + {\lambda }^{\frac{5}{2}} + q^4\,{\lambda }^{\frac{5}{2}} - q^{\frac{5}{2}}\,{\lambda }^3 \right. \nonumber \\
&&  + 
       q^6\,{\sqrt{\lambda }}\,\left( 1 + \lambda  \right)  + {\sqrt{q}}\,{\lambda }^2\,\left( 1 + \lambda  \right)  + 
       q^3\,{\sqrt{\lambda }}\,{\left( 1 + \lambda  \right) }^2 - q^5\,{\sqrt{\lambda }}\,{\left( 1 + \lambda  \right) }^2 \nonumber \\
&&  - 
       q^{\frac{3}{2}}\,\lambda \,{\left( 1 + \lambda  \right) }^2 - q^{\frac{9}{2}}\,\lambda \,\left( 2 + \lambda  \right)  - 
       q^2\,{\lambda }^{\frac{3}{2}}\,\left( 1 + 2\,\lambda  \right)  - q^{\frac{11}{2}}\,\left( -1 + \lambda  + {\lambda }^2 \right) \nonumber \\
&& \left.\left. + 
       q^{\frac{13}{2}}\,\left( 1 + \lambda  + {\lambda }^2 \right)  + 
       q^{\frac{7}{2}}\,\left( 1 + \lambda  + 3\,{\lambda }^2 + {\lambda }^3 \right)  \right)^2 \right]
\end{eqnarray} 
\subsection{Hopf link with framing $p_1$ on the first strand and $p_2$ on the second}
\begin{equation}
g_{\one,\one} = \frac{{\left( -1 \right) }^{p_1 + p_2} {\lambda^p}
\,\left( -1 + q \right) \,\left( -1 + \lambda  \right) }
  {{\sqrt{q}}\,{\sqrt{\lambda }}}
\end{equation}
\begin{equation}
g_{\twohor,\one} = \frac{{\left( -1 \right) }^{p_2}\,\left( -1 + \lambda  \right) \,
    \left( -q^{\frac{1}{2} + p_1} - 2\,q^{\frac{3}{2}}\,\left( -1 + \lambda  \right)  + 
      q^{\frac{3}{2} + p_1}\,\left( -1 + \lambda  \right)  + q^{\frac{5}{2} + p_1}\,\lambda  \right) }
    {q^{\frac{3}{2}}\,\lambda }
\end{equation}
\begin{equation}
g_{\twover,\one} = -\left( \frac{{\left( -1 \right) }^{p_2}\,q^{-1 -p_1}\,\left( -1 + \lambda  \right) \, \left( q + q^2 + 2\,q^{1 + p_1}\,\left( -1 + \lambda  \right)  - \lambda  - q\,\lambda  \right) }{\lambda }
    \right)
\end{equation}
These reformulated polynomials obey Marino's conjecture (\ref {horg}).
 
\end{document}